\documentclass[fleqn,usenatbib]{mnras}

\usepackage[T1]{fontenc}
\DeclareRobustCommand{\VAN}[3]{#2}
\let\VANthebibliography\thebibliography
\def\thebibliography{\DeclareRobustCommand{\VAN}[3]{##3}\VANthebibliography}
\usepackage{graphicx}
\usepackage{amsmath, amssymb}
\usepackage{threeparttable,multirow}
\usepackage{newtxtext,newtxmath}
\usepackage{color,soul}

\def\arcsec{$^{\prime\prime}$}
\def\kms{\,km\,s$^{-1}$}
\def\stars{{\!\ast}}
\def\gas{{\rm gas}}
\def\aper{{\rm ap}}

\def\re{$R_{\rm e}$}
\def\rre{\re/$R_{\rm PSF}$}
\def\rvg{$\nabla V_\gas$}

\def\dsig{$\Delta\sigma$}
\def\dxtwo{$\Delta\sigma_{X_{\rm two}}$}
\def\bpt{$C_{\rm AGN}$}

\title[Gas and stellar velocity dispersions]{The SAMI Galaxy Survey: the difference between ionised gas and stellar velocity dispersions}

\author[S.~Oh et~al.]{Sree~Oh,$^{1,2}$\thanks{E-mail: sree.oh@anu.edu.au} 
Matthew~Colless,$^{1,2}$
Francesco D'Eugenio,$^{3,4}$
Scott M. Croom,$^{2,5}$
\newauthor
Luca Cortese,$^{2,6}$
Brent Groves,$^{2,6}$
Lisa J. Kewley,$^{1,2}$
Jesse van de Sande,$^{2,5}$
Henry Zovaro,$^{1,2}$
\newauthor
Mathew R. Varidel,$^{7}$
Stefania Barsanti,$^{1,2}$
Sarah Brough,$^{2,8}$
Julia J. Bryant,$^{2,5}$
\newauthor
Sarah Casura,$^9$
Jon S. Lawrence,$^{10}$
Nuria P. F. Lorente,$^{10}$
Anne M. Medling, $^{2,11}$
\newauthor
Matt S. Owers,$^{2,12,13}$
Sukyoung K. Yi,$^{14}$
\\ ~ \\
$^{1}$Research School of Astronomy and Astrophysics, Australian National University, Canberra, ACT 2611, Australia\\
$^{2}$ARC Centre of Excellence for All Sky Astrophysics in 3 Dimensions (ASTRO 3D), Australia\\
$^{3}$Sterrenkundig Observatorium, Universiteit Gent, Krijgslaan 281 S9, B-9000 Gent, Belgium\\
$^{4}$Cavendish Laboratory and Kavli Institute for Cosmology, University of Cambridge, Madingley Rise, Cambridge, CB3 0HA, United Kingdom\\
$^{5}$Sydney Institute for Astronomy (SIfA), School of Physics, The University of Sydney, NSW 2006, Australia\\
$^{6}$International Centre for Radio Astronomy Research, The University of Western Australia, Crawley WA 6009, Australia\\
$^{7}$Brain and Mind Centre, The University of Sydney, 94 Mallett St, Camperdown, Sydney, NSW 2050, Australia\\
$^{8}$School of Physics, University of New South Wales, NSW 2052, Australia\\
$^{9}$Hamburger Sternwarte, Universitaet Hamburg, Gojenbergsweg 112, D-21029 Hamburg, Germany\\
$^{10}$Australian Astronomical Optics, Macquarie University, 105 Delhi Rd, North Ryde, NSW 2113, Australia\\
$^{11}$Ritter Astrophysical Research Center, University of Toledo, Toledo, OH 43606, USA\\
$^{12}$Department of Physics and Astronomy, Macquarie University, NSW 2109, Australia\\
$^{13}$Astronomy, Astrophysics and Astrophotonics Research Centre, Macquarie University, Sydney, NSW 2109, Australia\\
$^{14}$Department of Astronomy and Yonsei University Observatory, Yonsei University, Seoul 03722, Republic of Korea
}
\date{Accepted XXX. Received YYY; in original form ZZZ}

\begin{document}
\label{firstpage}
\pagerange{\pageref{firstpage}--\pageref{lastpage}}\pubyear{2021}
\maketitle

\begin{abstract}
We investigate the mean locally-measured velocity dispersions of ionised gas ($\sigma_\gas$) and stars ($\sigma_\stars$) for 1090 galaxies with stellar masses $\log\,(M_\stars/M_{\odot}) \geq 9.5$ from the SAMI Galaxy Survey. For star-forming galaxies, $\sigma_\stars$ tends to be larger than $\sigma_\gas$, suggesting that stars are in general dynamically hotter than the ionised gas (asymmetric drift). The difference between $\sigma_\gas$ and $\sigma_\stars$ (\dsig) correlates with various galaxy properties. We establish that the strongest correlation of \dsig\ is with beam smearing, which inflates $\sigma_\gas$ more than $\sigma_\stars$, introducing a dependence of \dsig\ on both the effective radius relative to the point spread function and velocity gradients. The second-strongest correlation is with the contribution of active galactic nuclei (AGN) (or evolved stars) to the ionised gas emission, implying the gas velocity dispersion is strongly affected by the power source. In contrast, using the velocity dispersion measured from integrated spectra ($\sigma_\aper$) results in less correlation between the aperture-based \dsig\ (\dsig$_\aper$) and the power source. This suggests that the AGN (or old stars) dynamically heat the gas without causing significant deviations from dynamical equilibrium. Although the variation of \dsig$_\aper$ is much smaller than that of \dsig, a correlation between \dsig$_\aper$ and gas velocity gradient is still detected, implying there is a small bias in dynamical masses derived from stellar and ionised gas velocity dispersions.
\end{abstract}

\begin{keywords}
galaxies: kinematics and dynamics -- galaxies: fundamental parameters -- galaxies: evolution -- galaxies: stellar content -- galaxies: active -- galaxies: structure
\end{keywords}

\section{Introduction}
\label{sec:intro}

The velocity dispersion in galaxies (especially that measured from aperture spectra) is often used as a proxy for the dynamical mass. It has been an important issue whether gas and stellar velocity dispersions trace the same dynamical potential, considering only one of them is measurable in many circumstances. Previous studies have reported a good agreement between gas and stellar velocity dispersions measured using aperture spectra, and justified the use of both gas and stellar velocity dispersions to trace the dynamical mass (e.g.\ Nelson \& Whittle 1996; Colina et~al.\ 2005; Greene \& Ho 2005; Chen, Hao, \& Wang 2008; Gilhuly, Courteau, \& Sa\'nchez et~al.\ 2019; but see also Ho 2009). However, the finding that both gas and stars trace a similar dynamical potential does not always indicate they have identical kinematics. The aperture velocity dispersion integrates both rotations and line-of-sight dispersions within an aperture, and is therefore a good approximation to the dynamical potential but misses the relative contribution of these. Integral-field spectroscopy (IFS) enables the local line-of-sight velocity dispersions to be separated from rotation, and thus potentially reveals physical processes tightly associated with gas and stars.

Emission-line samples for a reliable measure of gas kinematics are biased toward star-forming galaxies where the ionised-gas emission usually comes from H{\small II} regions confined to a thin disc plane displaying rotationally-supported kinematics (e.g.\ Barro et~al.\ 2016). Young stars may have similar kinematics to star-forming gas, but the increasing amount of pressure support in old stars leads to higher velocity dispersions in stars than in star-forming gas (asymmetric drift). On the other hand, gas is sensitive to non-gravitational motions such as winds and radiation driving from young stars and active galactic nuclei (AGN), in/outflows, and turbulence that can lead to enhanced gas velocity dispersions (e.g.\ Dib et~al.\ 2006; Agertz et~al.\ 2009; Tamburro et~al.\ 2009; Aumer et~al.\ 2010; Wisnioski et~al.\ 2011; Rich, Kewley, \& Dopita 2015; Ho et~al.\ 2014; Woo, Son \& Bae 2017; Zhou et~al.\ 2017; Orr et~al.\ 2020). Stellar velocity dispersions are weakly correlated to such non-gravitational processes, and therefore, they can serve as a reference point to quantify the impact of feedback processes on the gas velocity dispersion. 

Different responses of gas and stars to the gravitational and non-gravitational processes suggest their kinematics also have different correlations to galaxy properties. Ho (2009) compared stellar and gas velocity dispersions measured using the spectra integrated through an aperture of 2\arcsec$\times$4.1\arcsec~from Paloma long-slit data (Ho et~al.\ 1997) and reported evidence of AGN feedback. Barat et~al.\ (2019) found that galaxy morphology correlates to the difference between locally-measured gas and stellar velocity dispersions. Foster et~al.\ (in prep.) also find a link between the stellar population and the dynamical coupling of ionised gas and stars. However, the difference between ionised gas and stars in locally-measured velocity dispersions with various galaxy properties has not been actively explored to reveal the physical processes directly involved. 

This work quantifies the difference between gas and stellar velocity dispersions using a large sample of IFS galaxies from the Sydney-AAO Multi-object Integral-field spectroscopy (SAMI) Galaxy Survey. We examine the relation between the difference in gas and stellar velocity dispersions and galaxy properties such as stellar mass, size, bulge fraction, stellar populations, star-formation rate, and AGN. We identify parameters suspected of having a causal connection to the differences in velocity dispersions and suggest possible scenarios. We also revisit the aperture velocity dispersions and discuss them as dynamical mass estimates.

This paper is outlined as follows. The data and sample are described in Section~\ref{sec:data}. In Section 3, we outline the method for measuring stellar and gas velocity dispersions. In Section 4 we present our results identifying the primary galaxy parameters involved in the gas and stellar velocity dispersions. In Section 5 we discuss possible explanations for the different gas and stellar velocity dispersions. We draw our conclusion in Section 6. Throughout the paper, we assume a standard $\Lambda$CDM cosmology with $\Omega_m = 0.3$, $\Omega_{\Lambda} = 0.7$, and $H_0 = 70$\kms~Mpc$^{-1}$.
 
\section{Data and sample}
\label{sec:data}

\subsection{The SAMI Galaxy Survey}
\label{sec:sami}

SAMI is a multi-object fibre integral field system (Croom et~al.\ 2012), composed of 13 15\arcsec-diameter hexabundles, each containing 61 1\arcsec.6-diameter optical fibres (Bland-Hawthorn et~al.\ 2011; Bryant et~al.\ 2011, 2014). The hexabundles are fed into the AAOmega dual-arm spectrograph mounted on the 3.9-metre Anglo-Australian Telescope (Sharp et~al.\ 2006). The 580V and 1000R gratings are used for blue (3750--5750\,\AA) and red (6300--7400\,\AA) arms, yielding the spectral resolution of R = 1808 and R = 4304 respectively. The spectral resolutions in the blue and red arms are equivalent to an effective velocity dispersion of 70.4 and 29.6\kms~respectively (van de Sande et~al.\ 2017b; Scott et~al.\ 2018).

The final data release of the SAMI Galaxy Survey includes more than 3000 unique galaxy cubes at redshifts $0.04 < z < 0.095$ (Bryant et~al.\ 2015; Croom et~al.\ 2021). The majority of SAMI targets are drawn from the three equatorial regions of the Galaxy And Mass Assembly (GAMA; Driver et~al.\ 2011) survey. In addition, eight cluster regions have been observed to complete the environmental matrix (Owers et~al.\ 2017). 

\subsection{Galaxy parameters}
\label{sec:par}

In this section, we present a brief description and indicate the primary reference for each galaxy parameter used in this study. The stellar masses ($M_\stars/M_{\odot}$) have been derived using $i$-band magnitudes and $k$-corrected $g - i$ colours (Taylor et~al.\ 2011; Bryant et~al.\ 2015). The photometric parameters of position angle, ellipticity, and half-light radius (\re) are derived from the Sloan Digital Sky Survey (SDSS; York et~al.\ 2000) and the VLT Survey Telescope (VST) ATLAS Survey (Shanks et~al.\ 2013) imaging using the Multi Gaussian Expansion algorithm (MGE, Emsellem et~al.\ 1994), as implemented in the {\sc mgefit} package (Cappellari 2002). A full description of the MGE measurements is in D'Eugenio et~al.\ (2021). The bulge-to-total ratio (B/T) have been measured employing ProFit, a routine for Bayesian two-dimensional galaxy profile modelling (Robotham et~al.\ 2017), to $r$-band images from the Kilo-Degree Survey (KiDS; de Jong et~al.\ 2017), the VLT/ATLAS survey, and the SDSS (Casura et~al.\ subm.; Barsanti et~al.\ 2021). We derived single stellar population (SSP) equivalent measurements of the light-weighted stellar age and metallicity ([Z/H]) integrating within an elliptical aperture of 1 \re~(Scott et~al.\ 2017, 2018). We also tested the age and metallicity from the full spectral fitting method described in Barone et~al.\ (2018, 2020) and derived the same conclusion. We fit for the H$\beta$, [O\,{\small III}] $\lambda$5007, H$\alpha$, [N\,{\small II}] $\lambda$6583 emission lines with a Gaussian profile using the fitting code {\sc lzifu} (Ho et~al.\ 2016a). The emission-line fluxes measured using a single Gaussian fit are used for emission-line diagnostics in Section~\ref{sec:bpt}. The star-formation rate (SFR) has been derived using the dust-corrected H$\alpha$ emission flux from star-forming spaxels, assuming an intrinsic Balmer decrement of H$\alpha$/H$\beta$ = 2.86 (e.g.\ Medling et~al.\ 2018). 

\subsection{Quantifying emission-line diagnostics}
\label{sec:bpt}

\begin{figure}
\centering
\centering\includegraphics[width=\columnwidth]{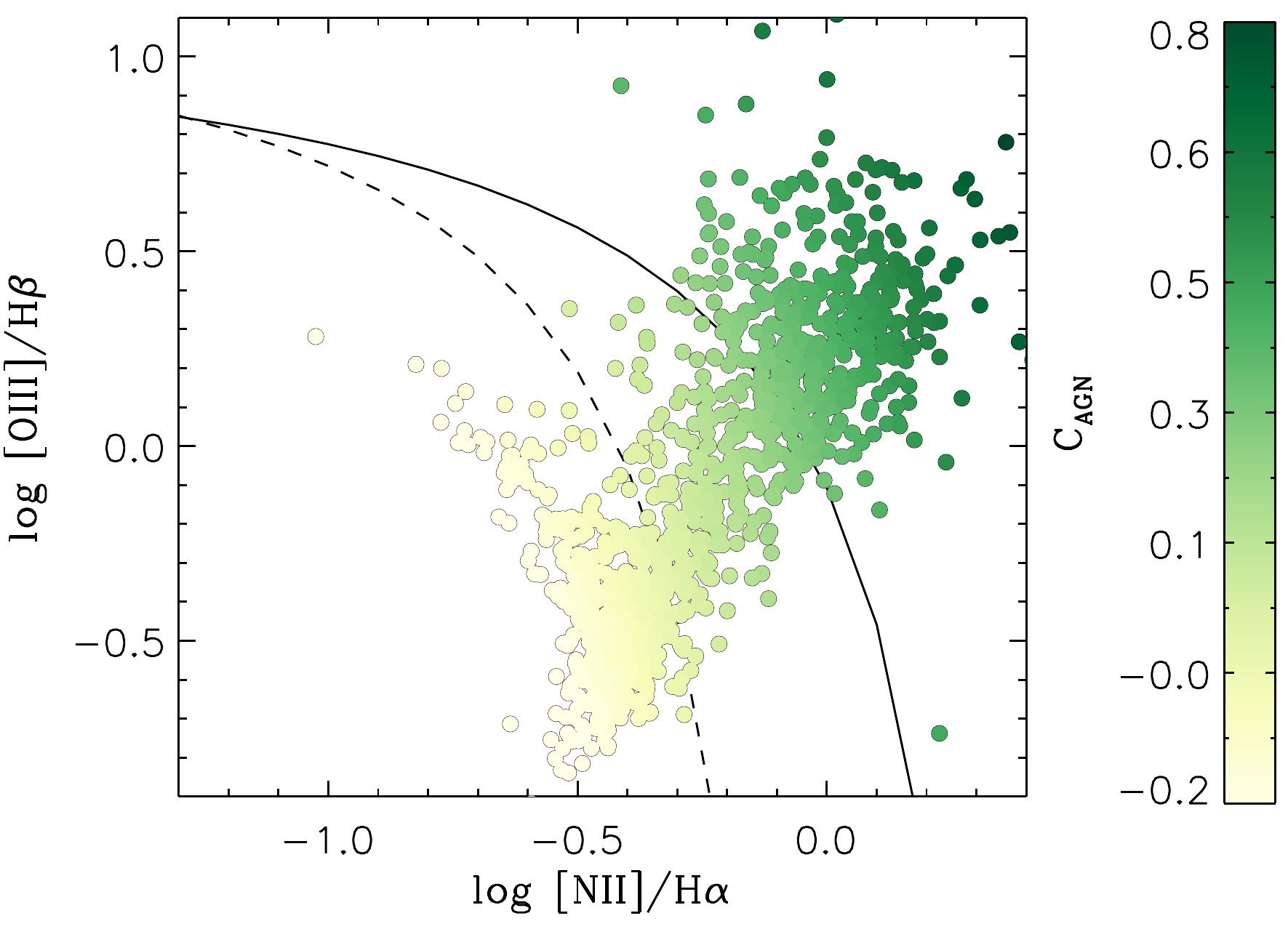}
\caption{Emission-line diagnostics for the 1090 sample galaxies. We used emission line fluxes integrated within 1 \re. The solid and dashed lines are, respectively, the demarcation lines of Kewley et~al.\ (2001) and Kauffmann et~al.\ (2003), separating star-forming galaxies and AGN. We measure the contribution of AGN (\bpt) as the orthogonal departure (i.e. the smallest distance) in log scale from the dashed line (log [O\,{\small III}]/H$\beta$ = 0.61 / (log [N\,{\small II}]/H$\alpha$-0.05)+1.3). Galaxies are colour-coded by \bpt.}
\label{bpt}
\end{figure}

We used emission-line diagnostics to quantify the relative contribution of ionised gas powering sources as one of the key characteristics of galaxies. Optical emission-line diagnostics, so-called BPT diagnostics (Baldwin, Phillips \& Telervich 1981), employ line ratios between Balmer and forbidden lines to empirically separate galaxies whose emission lines are dominated by AGN from those powered by star-forming populations (e.g.\ Kewley et~al.\ 2006). In Figure~\ref{bpt}, we present emission-line diagnostics for the sample using the ratio of integrated H$\alpha$, H$\beta$, [N\,{\small II}], and [O\,{\small III}] line fluxes within 1\,\re\footnote{The use of the mean of the ratio of the fluxes per spaxel within 1\,\re~for the emission-line diagnostics does not change the results of the study.}. Emission-line diagnostics indicate the relative contributions of power sources for the ionised gas. We define a \bpt\ parameter as the orthogonal departure in log scale from the empirical demarcation line provided by Kauffmann et~al.\ (2003) (log [O\,{\small III}]/H$\beta$ = 0.61 / (log [N\,{\small II}]/H$\alpha$-0.05)+1.3; dashed line in Figure~\ref{bpt}), which has been widely used to quantify the AGN (or star-formation) contribution or the mixing between star formation and AGN to the ionised gas emission (e.g.\ Kauffmann \& Heckman 2009; Jones et~al.\ 2016; Thomas et~al.\ 2018). Star-forming galaxies and AGNs have, respectively, low and high values of \bpt, by definition. However, we also note that \bpt\ may not necessarily indicate the contribution of AGN when there are alternative ionisation sources (e.g. old evolved stars). We discuss ionisation sources for AGN-like emission and their implications in Section~\ref{sec:liner}.
  
\subsection{Sampling criteria}
\begin{figure}
\centering
\centering\includegraphics[width=\columnwidth]{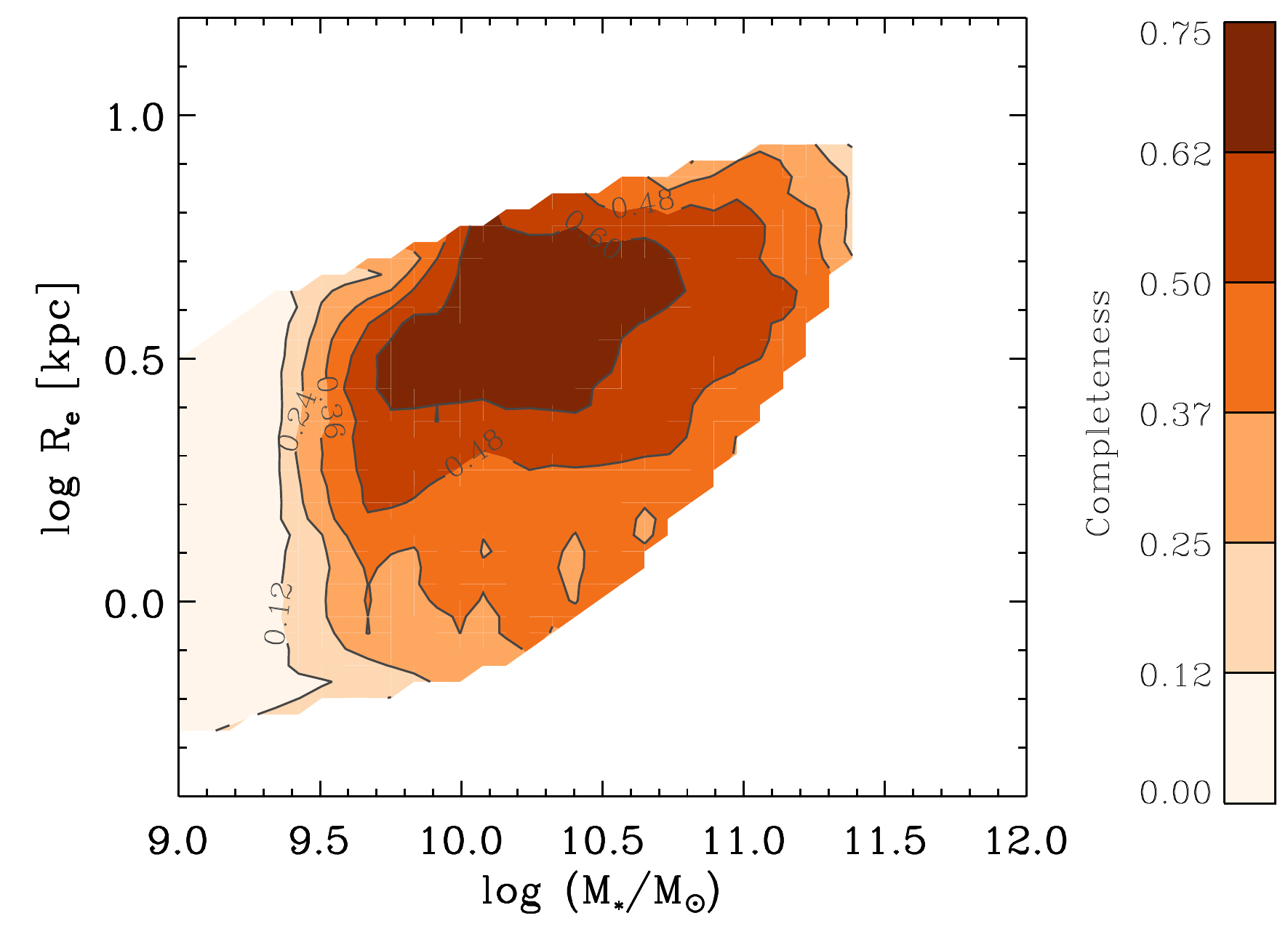}
\caption{Sampling completeness derived as the fraction of the final sample relative to the parent SAMI galaxies at each point.}
\label{sam}
\end{figure}  

We used the final SAMI data release which has been described in the Data Release 3 publication (Croom et~al.\ 2021). The SAMI sample, including 3000 galaxies, is volume-limited (Bryant et~al.\ 2015; Owers et~al.\ 2017). We first selected 2162 galaxies whose stellar mass is greater than 10$^{9.5} M_{\odot}$ to ensure unbiased stellar and gas velocity dispersions (given the spectral resolution limit of SAMI and low signal-to-noise (S/N) in the low mass galaxies) and to avoid large volume correction factors \footnote{The choice of the mass limit for unbiased velocity dispersion measurements has been discussed in, e.g., van de Sande et~al.\ (2017b) and Barat et~al.\ (2019, 2020). We also note that applying a conservative mass cut of 10$^{10} M_{\odot}$ does not change the results of this study.}. We then excluded 1072 galaxies because more than 10\% of spaxels within \re~have lower signal-to-noise than 3~\AA$^{-1}$ either in the continuum or H$\alpha$ emission flux, which results in a final sample of 1090 galaxies. Although small galaxies relative to the point spread function (PSF) size are significantly affected by beam smearing, we do not exclude them from the sample considering the PSF size is not the only factor contributing to beam smearing (e.g.\ Glazebrook 2013; Bassett et~al.\ 2017; Varidel et~al.\ 2019). We discuss the impact of beam smearing on the velocity dispersion in Section~\ref{sec:beam}. 

Sampling completeness was measured as the fraction of the final sample relative to the parent SAMI sample at each point in the mass--size plane (Figure~\ref{sam}). Our final sample is biased toward emission-line galaxies due to the S/N cut applied to H$\alpha$ emission flux. Accordingly, the sampling completeness is high in a specific area in the mass--size plane where we usually find star-forming galaxies. The impact of sampling bias on the result is discussed in Section~\ref{sec:xthree}.

\section{Measurement of velocity dispersions}
\label{method}

In this section, we briefly introduce the methods to derive gas and stellar kinematics, which are the same as the standard methods that the SAMI team used (Croom et~al.\ 2021); we explain in the following where our methods differ from DR3.
 
\subsection{Stellar velocity dispersion}
\label{ppxf}

The stellar velocity dispersion has been estimated using the penalized pixel fitting ({\sc pPXF}; Cappellari \& Emsellem 2004; Cappellari 2017) method. We first prepared SAMI spectra combining blue (3750--5750\,\AA) and red (6300--7400\,\AA) spectra to have the constant spectral resolution 2.65\,\AA~over the full wavelength range. The spectra also have been de-redshifted and re-binned on to a logarithmic wavelength scale. As templates for the full spectral fitting, we used the MILES stellar library (S\'anchez-Bl\'azquez et~al.\ 2006), consisting of 985 stellar spectra, which have been convolved to the same resolution as the SAMI spectra (2.65\,\AA) from their original resolution of 2.51\,\AA\ (Falcon-Barroso et~al.\ 2011). To avoid degeneracies and to ensure accurate kinematics in low S/N spaxels, we pre-select the templates from high-S/N (a minimum of 25\,\AA$^{-1}$) annular-binned spectra, which are generated from five equally-spaced elliptical annuli following the light distribution of a galaxy. See Croom et~al. (2021) and van de Sande et~al. (2017b) for more details on the spaxel binning scheme of SAMI. We used a 12\textsuperscript{th}-order additive Legendre polynomial and fit a Gaussian line-of-sight velocity distribution to extract the rotation velocity and velocity dispersion. We ran {\sc pPXF} three times: the first run improves the estimate of input noise, the second run determines the wavelength pixels to use, and the third run measures stellar kinematics with improved noise estimates, only including `good' wavelength pixels. See van de Sande et~al.\ (2017b) for more details on the stellar kinematics of SAMI.

\subsection{Gas velocity dispersion}

The gas velocity dispersion has been estimated using the emission line fitting code {\sc lzifu} (Ho et~al.\ 2016a). We first subtracted the continuum derived using {\sc pPXF}, as described in Owers et~al.\ (2019). We then fit [O\,{\small II}] $\lambda$3727, 3729, H$\beta$, [O\,{\small III}] $\lambda$5007, H$\alpha$, [N\,{\small II}] $\lambda$6583, [S\,{\small II}] $\lambda$6716, and [S\,{\small II}] $\lambda$6731 emission lines simultaneously to estimate the mean rotation velocity and velocity dispersion. We used the result from single-component Gaussian fits.

\subsection{Mean gas and stellar velocity dispersion}
\label{global}

\begin{figure}
\centering
\centering\includegraphics[width=\columnwidth]{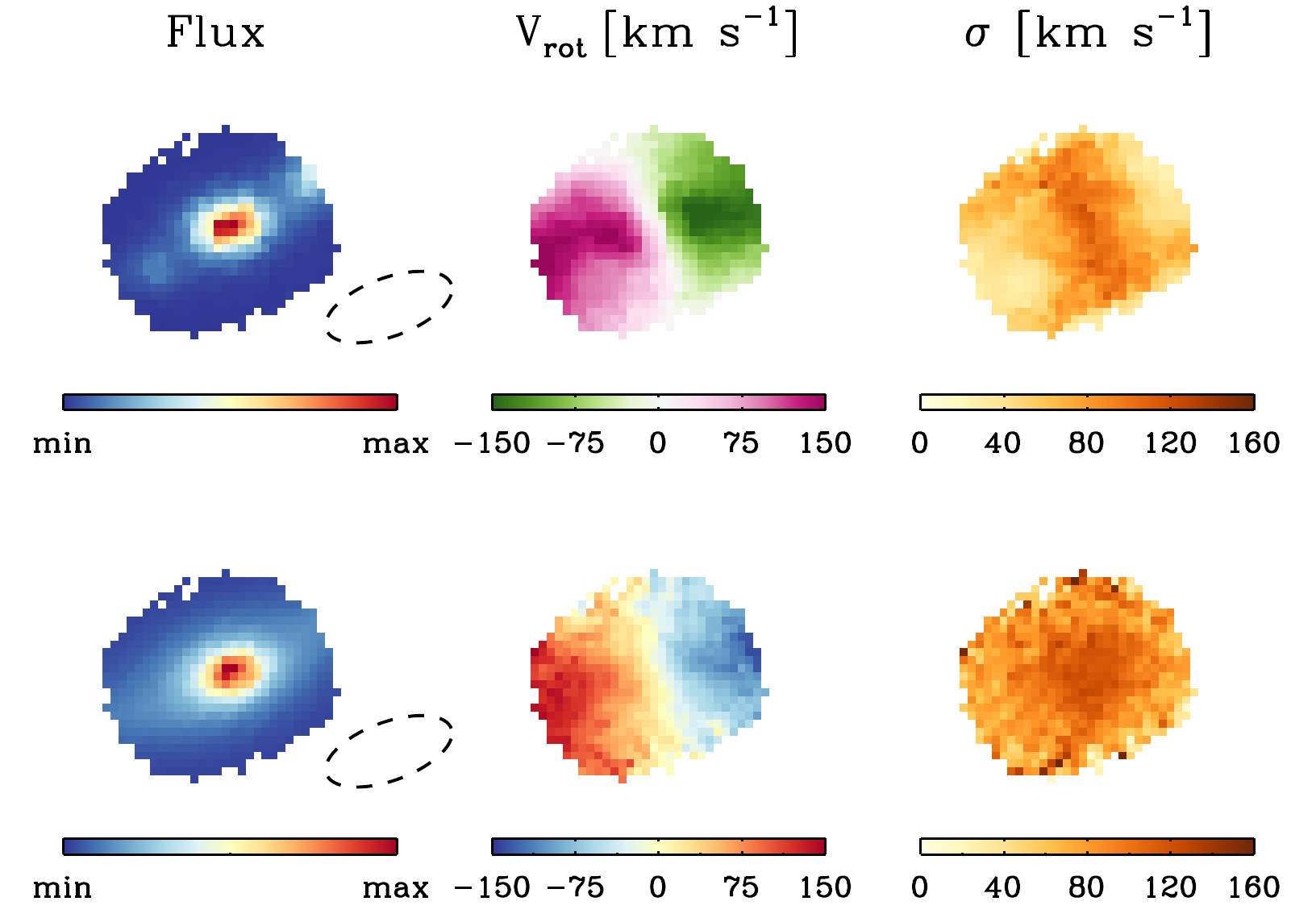}
\caption{Example gas (top) and stellar (bottom) kinematics for SAMI 105597. The first column shows H$\alpha$ flux (for gas) and continuum flux (for stars). The second and third columns present, respectively, the rotation velocity and velocity dispersions. The ellipse indicates 1 \re~in which $\sigma_\stars$ and $\sigma_\gas$ have been integrated. We only show spaxels whose S/N is greater than 3\,\AA$^{-1}$.}
\label{kinmap}
\end{figure}

We present an example of spatially-resolved gas and stellar kinematics in Figure~\ref{kinmap}. We measure the average of the gas and stellar velocity dispersions ($\sigma_\stars$ and $\sigma_\gas$) within 1\,\re\ as
\begin{equation} \sigma_\stars^2 \equiv \frac{\sum_i^{N_{\rm spaxel}} \sigma_{\stars,i}^2}{N_{\rm spaxel}} ~, \end{equation}
\begin{equation} \sigma_\gas^2 \equiv \frac{\sum_i^{N_{\rm spaxel}} \sigma_{\gas,i}^2}{N_{\rm spaxel}} ~, \end{equation}
where $\sigma_{\stars,i}$ and $\sigma_{\gas,i}$ are, respectively, the stellar and gas velocity dispersions measured in the $i$\textsuperscript{th} spaxel. $N_{\rm spaxel}$ is the number of spaxels used within \re. $\sigma_\stars$ and $\sigma_\gas$ are measured using the same spaxels where both the continuum flux-based and H$\alpha$ flux-based S/Ns are greater than 3\,\AA$^{-1}$.
The difference between the gas and stellar velocity dispersions (\dsig) is defined as
\begin{equation} \Delta\sigma \equiv \log\sigma_\gas - \log\sigma_\stars. \end{equation}

Note that both $\sigma_\stars$ and $\sigma_\gas$ are unweighted sums; we did not apply flux weighting to avoid the large uncertainty in the dust correction for galaxies with AGN-like emission and bias in $\Delta\sigma$ according to the different light distributions between the continuum and emission fluxes. We compared the results if we approached this using a luminosity-weighted sum and found qualitatively consistent results, and stay with the unweighted results for simplicity. We also tested another approach calculating $\Delta\sigma$ as the median of the ratio of $\sigma_{\rm gas,i}/\sigma_{\rm *,i}$ within $R_{\rm e}$ and found qualitatively consistent results.

\subsection{Velocity gradients}

At any spaxel $(x,y)$ on the velocity map, we measured the magnitude of the velocity gradient ($\nabla V$) from the line-of-sight velocity of neighbouring spaxels ($V_\mathrm{los}$) following Varidel et~al.\ (2016):
\begin{equation} \nabla V(x,y) \equiv \frac{1}{2 W_{\rm pix}}\sqrt{\begin{aligned} &[V_{\rm los}(x+1,y)-V_{\rm los}(x-1,y)]^2 \\
     & + [V_{\rm los}(x,y+1)-V_{\rm los}(x,y-1)]^2
    \end{aligned}
    }~, \end{equation}
where $W_{\rm pix}$ is the size of the spaxel that is 0.5\arcsec\ for all the sample galaxies. $V_{\rm los}$ is not corrected for inclination because we mainly consider $\nabla V$ as an indicator for an observational limitation (i.e.\ beam smearing) rather than a physical property of galaxies (i.e.\ intrinsic rotation). The mean gas velocity gradients ($\nabla V_\gas$) have been calculated as the unweighted mean within \re, as in Equations~1 \&~2.

\section{Results}
\label{sec:res}

\subsection{Gas and stellar velocity dispersions}

\begin{figure}
\centering
\centering\includegraphics[width=\columnwidth]{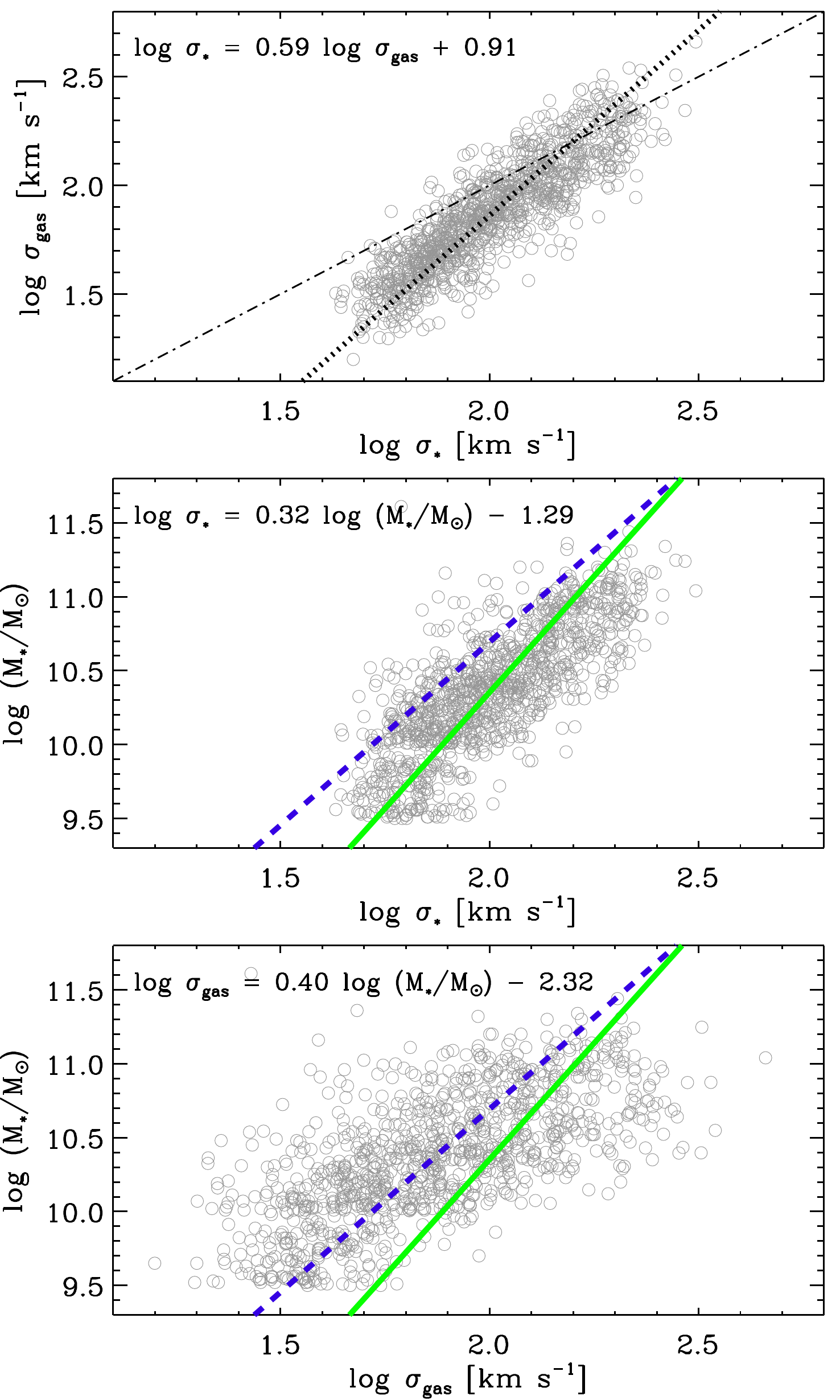}
\caption{A comparison between gas and stellar velocity dispersions. In the top panel, the dotted line is a linear fit, and the dash-dot line is the one-to-one line. The solid and dashed lines are, respectively, linear fits to the stellar and gas $M_{*}$--$\sigma$ relations. Note that the typical errors of $\log\sigma_\stars$ and $\log\sigma_\gas$ are, respectively, 0.006 and 0.005 dex.}
\label{comp}
\end{figure}

In Figure~\ref{comp}, we show that the gas velocity dispersion ($\sigma_\gas$) is in general smaller than the stellar velocity dispersion ($\sigma_\stars$), confirming that the gas is dynamically colder than the stars. However, we also find that $\sigma_\gas$ is not always smaller than $\sigma_\stars$. While $\sigma_\gas$ is much smaller than $\sigma_\stars$ at lower stellar mass galaxies, the difference decreases as the stellar mass increases. In massive galaxies, $\sigma_\gas$ is comparable to or sometimes even larger than $\sigma_\stars$. Note that the typical errors in $\log\sigma_\stars$ and $\log\sigma_\gas$ are, respectively, 0.006\,dex and 0.005\,dex. Both $\sigma_\stars$ and $\sigma_\gas$ have positive correlations with stellar mass. However, the $M_\stars$--$\sigma_\gas$ relation has a shallower slope with a larger scatter than the stellar $M_\stars$--$\sigma_\stars$ relation. The difference between gas and stellar scaling relations has also been reported in Cortese et~al.\ (2014) and Barat et~al.\ (2019). Different slopes in the $M_\stars$--$\sigma_\stars$ and the $M_\stars$--$\sigma_\gas$ relations support the hypothesis that they trace different locations within the gravitational potential. 

\begin{figure*}
%\centering\includegraphics[width=\textwidth]{fig5.eps}
\centering\includegraphics[width=\textwidth]{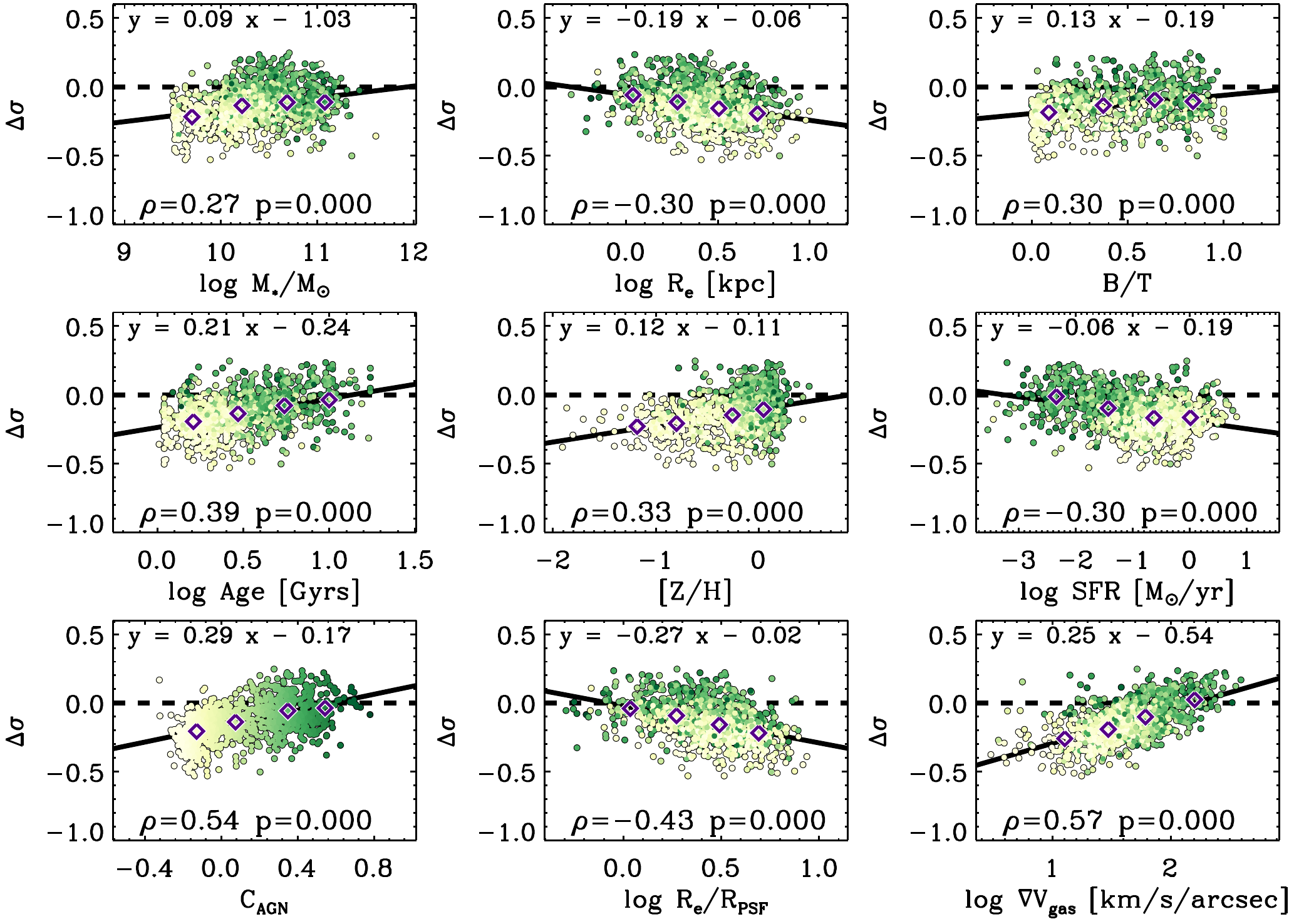}
\caption{The dependence of \dsig\ on various galaxy parameters. We present a linear fit to each relation on the top and Spearman's coefficient $\rho$ and $p$-value at the bottom of each panel. \dsig\ correlates with various galaxy parameters, with $|\rho|=0.27$--0.57. The data are colour-coded by \bpt\ (itself shown in the bottom left panel). Diamonds indicate the median \dsig\ for each bin. The typical error of \dsig\ is around 0.01\,dex.}
\label{sigdiff}
\end{figure*}

In Figure~\ref{sigdiff}, we present the dependence of \dsig\ on various galaxy parameters representing structures ($M_\stars$, $R_{\rm e}$, and B/T), populations (Age, [Z/H], SFR, and \bpt), and observational limitations (\rre\ and \rvg). To quantify the strength and significance of each correlation, we derived the Spearman correlation coefficient $\rho$ for each correlation. We find \dsig\ is strongly correlated with log\,\rvg\ (with $\rho=0.57$) and \bpt\ (with $\rho=0.54$). We also find a strong correlation between \dsig\ and log\, ($\nabla V_{\rm gas}/\nabla V_{\rm *}$) with $\rho=0.52$ (Appendix~\ref{sec:app1}). Moderate correlations are also detected from the relations with the other parameters ($|\rho|=0.27$--0.43). We note that \dsig\ does not correlate with ellipticity or inclination ($\rho=0$ for both parameters), although they are not shown in the figure. It is not straightforward to identify primary parameters driving \dsig\ solely based on $\rho$ given that many of the galaxy parameters explored here are correlated. 

\subsection{Primary parameters driving \texorpdfstring{$\boldsymbol{\Delta\sigma}$}{}}
\label{sec:comb}

\subsubsection{One primary parameter}
\label{sec:xone}

To identify the mechanisms producing the observed difference \dsig, we first test the simple hypothesis that \dsig\ is driven by a single parameter $X$, and the other observed correlations are all due to secondary correlations with $X$. We calculated $\Delta\sigma_{\rm X}$ using a linear fit to the relation between the parameter X and \dsig. Then, we subtracted $\Delta\sigma_{\rm X}$ from \dsig\ to remove the dependence of \dsig\ on the parameter X and to see whether this correction using $\Delta\sigma_{\rm X}$ also removes the correlations between \dsig\ and the other galaxy parameters. For example, we define $\Delta\sigma_{\rm C_{AGN}} = 0.29$\bpt$-0.17$ from the linear fit to the relation between \dsig\ and \bpt\ (the bottom left panel in Figure~\ref{sigdiff}). Then, we calculate $\rho$ for the relations between the corrected \dsig\ (\dsig\ - $\Delta\sigma_{\rm C_{AGN}}$) and the galaxy parameters. Zero values for $\rho$ are expected when the connection between \dsig\ and \bpt\ fully explains the correlations between \dsig\ and the other parameters.

We repeated the correction of \dsig\ with $\Delta\sigma_{\rm X}$ by changing the correction parameter $X$. $\rho$ from each correction is displayed in Table~\ref{tab:xone}. We still find significant residual correlations ($\rho \sim 0.2$--0.5) when applying $\Delta\sigma_{\log(M_\stars/M_{\odot})}$, $\Delta\sigma_{\rm B/T}$, $\Delta\sigma_{\rm log Age}$, $\Delta\sigma_{\rm [Z/H]}$, and $\Delta\sigma_{\rm log SFR}$, suggesting these parameters do not explain the dependence of \dsig\ on the other parameters. The correction using $\Delta\sigma_{\rm C_{AGN}}$ removes the dependence of \dsig\ on many other parameters, giving $\rho$ of almost zero, except for log\,\re, log\,(\rre), and log\,\rvg. Applying $\Delta\sigma_{R_{\rm e}}$ or $\Delta\sigma_{\log(R_{\rm e}/R_{\rm PSF})}$ does not eliminate the other correlations, but there are no other parameters that explain the relation between \dsig\ and log(\rre) (or log\,\re). Note that a residual correlation between log(\rre) and (\dsig$-\Delta\sigma_{\log R_{\rm e}}$) has been detected ($\rho = -0.23$). However, applying $\Delta\sigma_{\log (R_{\rm e}/R_{\rm PSF})}$ fully explains the correlation between log\,$R_{\rm e}$ and \dsig\ ($\rho = 0.01$). Therefore, log(\rre) seems to be the primary driver of the log\,$R_{\rm e}$--\dsig\ relation. The correction for log\,\rvg\ removes the dependence of \dsig\ on many other parameters, except for log(\rre) and \bpt. In conclusion, there is no single parameter that, alone, can explain all the correlations of \dsig\ with the observables considered here.	

\begin{table*}
\centering
\caption{Spearman correlation coefficients $\rho$ from the relation between (\dsig$-$$\Delta\sigma_X$) and galaxy parameters}
\begin{tabular}{lrrrrrrrrr}
\hline\hline
\dsig$_{\rm X}$ & log ($M_\stars$/$M_{\odot}$)& log R$_{\rm e}$& B/T & log Age & $[Z/H]$ & log SFR & \bpt & log (\rre) & log \rvg \\
\hline
no correction  & 0.27 & -0.30 & 0.30 & 0.39 & 0.33 & -0.29 & 0.54 & -0.43 & 0.57 \\
\hline
\dsig$_{\rm log (M_\stars/M_{\odot})}$  & \textbf{0.00} & -0.47 & 0.22 & 0.29 & 0.15 & -0.32 & 0.39 & -0.51 & 0.41 \\
\dsig$_{\rm log R_{\rm e}}$  & 0.45 & \textbf{-0.01} & 0.24 & 0.38 & 0.41 & -0.17 & 0.55 & -0.23 & 0.60 \\
\dsig$_{\rm B/T}$  & 0.17 & -0.26 & \textbf{0.02} & 0.27 & 0.24 & -0.22 & 0.42 & -0.35 & 0.45 \\
\dsig$_{\rm log Age}$ & 0.11 & -0.28 & 0.13 & \textbf{-0.01} & 0.18 & \textbf{-0.05} & 0.28 & -0.38 & 0.37 \\
\dsig$_{\rm [Z/H]}$ & \textbf{0.05} & -0.38 & 0.21 & 0.28 & \textbf{0.02} & -0.19 & 0.33 & -0.44 & 0.38 \\
\dsig$_{\rm log SFR}$  & 0.31 & -0.16 & 0.21 & 0.22 & 0.25 & \textbf{0.00} & 0.40 & -0.31 & 0.50 \\
\dsig$_{\rm C_{AGN}}$  & \textbf{-0.04} & -0.32 & \textbf{0.06} & \textbf{0.04} & -\textbf{0.02} & \textbf{-0.01} & \textbf{0.02} & -0.38 & 0.30 \\
\dsig$_{\rm log R_{\rm e}/R_{\rm PSF}}$  & 0.40 & \textbf{0.01} & 0.20 & 0.34 & 0.36 & -0.13 & 0.50 & \textbf{-0.01} & 0.53 \\
\dsig$_{\rm log \nabla V_\gas}$  & \textbf{-0.10} & -0.37 & \textbf{0.08} & 0.11 & \textbf{-0.01} & -0.16 & 0.23 & -0.38 & \textbf{0.03} \\
\hline
$\Delta\sigma_{\rm Xtwo,1}$  & -0.20 & -0.36 & \textbf{-0.00} & \textbf{-0.03} & -0.14 & \textbf{-0.03} & \textbf{0.00} & -0.37 & \textbf{-0.00} \\
$\Delta\sigma_{\rm Xtwo,2}$ & \textbf{0.03} & \textbf{-0.08} & \textbf{0.01} & \textbf{0.07} & \textbf{0.03} & \textbf{-0.02} & 0.22 & \textbf{-0.01} & \textbf{0.01} \\
$\Delta\sigma_{\rm Xtwo,3}$  & \textbf{0.09} & \textbf{-0.04} & \textbf{0.00} & \textbf{0.01} & \textbf{0.02} & 0.11 & \textbf{0.02} & \textbf{-0.02} & 0.29 \\
\hline
$\Delta\sigma_{\rm Xthree}$  & \textbf{-0.05} & \textbf{-0.08} & \textbf{-0.06} & \textbf{-0.05} & \textbf{-0.09} & \textbf{0.08} & \textbf{0.00} & \textbf{-0.00} & \textbf{-0.00} \\
\hline\hline
\multicolumn{10}{l}{\textbf{Note}. The correction factor $\Delta\sigma_{\rm X}$ has been derived using a linear fit to the relation between X and \dsig\ (Figure~\ref{sigdiff}).}\\
\multicolumn{10}{l}{It is highlighted in boldface when $|\rho| \leq 0.1$.}\\
\label{tab:xone}
\end{tabular}
\end{table*}

\subsubsection{Two primary parameters}

We now introduce a correction based on the assumption that two parameters independently contribute to the difference in the gas and stellar velocity dispersions. We define a correction parameter $X_{\rm two}$ = $X_1$ + $\alpha$ $X_2$ which is a linear combination of two parameters with $\alpha$ fixing the relative contribution of $X_1$ and $X_2$. We numerically estimate the $\alpha$ that makes the correction using \dxtwo, a linear fit to the relation between \dsig\ and $X_{\rm two}$, simultaneously eliminating the dependence of (\dsig$-$\dxtwo) on $X_1$ and $X_2$. Specifically, we calculated a sum of two correlation coefficients in quadrature, $\rho_{\rm two} = \sqrt{\rho_1^2 + \rho_2^2}$ with varying $\alpha$, where $\rho_1$ and $\rho_2$ are, respectively, calculated from the $X_1$--(\dsig$-$\dxtwo) and $X_2$--(\dsig$-$\dxtwo) relations. Then we selected $\alpha$ minimising $\rho_{\rm two}$. For example, we found $\alpha=-0.95$ when setting $X_1 = \log$\rvg\ and $X_2 = \log$(\rre), which yields $\rho_1 = -0.01$ for the relation between $X_1$ and (\dsig$-$\dxtwo) and $\rho_2 = 0.02$ for the relation between $X_2$ and (\dsig$-$\dxtwo). We applied this analysis combining key parameters (\rre, \rvg, and \bpt) that at least partially explain the other correlations and are unlikely to be explained by the other parameters. 

\begin{figure}
\centering
\centering\includegraphics[width=\columnwidth]{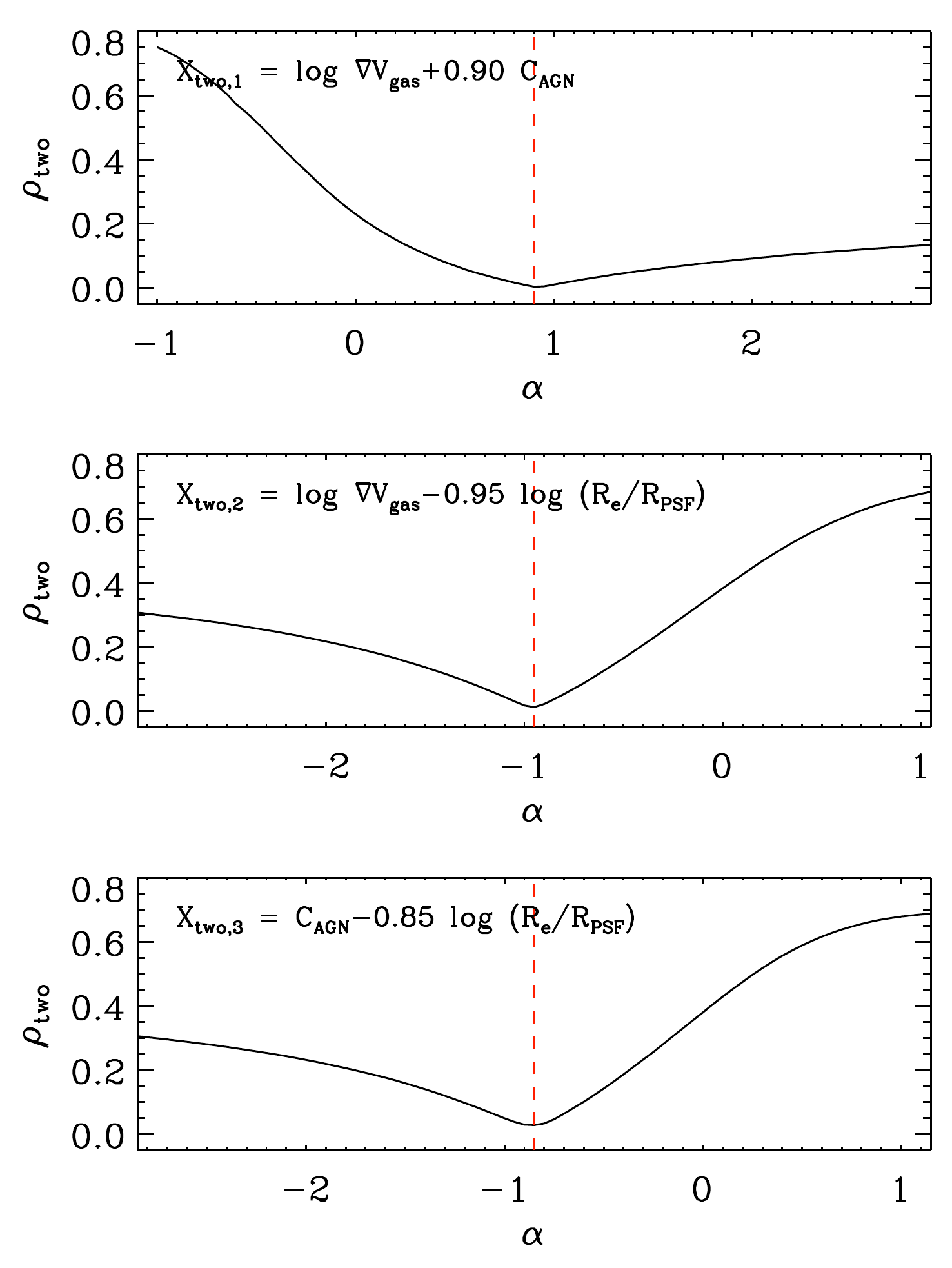}
\caption{The optimal $\alpha$ for $X_{\rm two} = X_1 + \alpha X_2$ is taken to be the one generating the minimum $\rho_{\rm two}= \sqrt{\rho_1^2 + \rho_2^2}$, where $\rho_1$ and $\rho_2$ are, respectively, calculated from
 the $X_1$--(\dsig$-$\dxtwo) and $X_2$--(\dsig$-$\dxtwo) relations. \dxtwo\ is measured using a linear fit to the relation between \dsig\ and $X_{\rm two}$ (see Equation~\ref{eq:xtwo}).  }
\label{rho_xtwo}
\end{figure}

In Figure~\ref{rho_xtwo}, we present the change in $\rho_{\rm two}$ when varying $\alpha$. The optimal $\alpha$ for each combination has been taken from the global minimum of $\rho_{\rm two}$. Finally, we defined $X_{\rm two}$ combining two key parameters with an optimal $\alpha$ as follows: 
\begin{align}
&X_{\rm two,1} = \log \nabla V_\gas + 0.90~C_{\rm AGN},\nonumber \\
&X_{\rm two,2} = \log \nabla V_\gas -0.95~\log (R_{\rm e}/R_{\rm PSF}),\\
&X_{\rm two,3} = C_{\rm AGN} - 0.85~\log (R_{\rm e}/R_{\rm PSF}).\nonumber
\end{align}

\begin{figure}
\centering
\centering\includegraphics[width=\columnwidth]{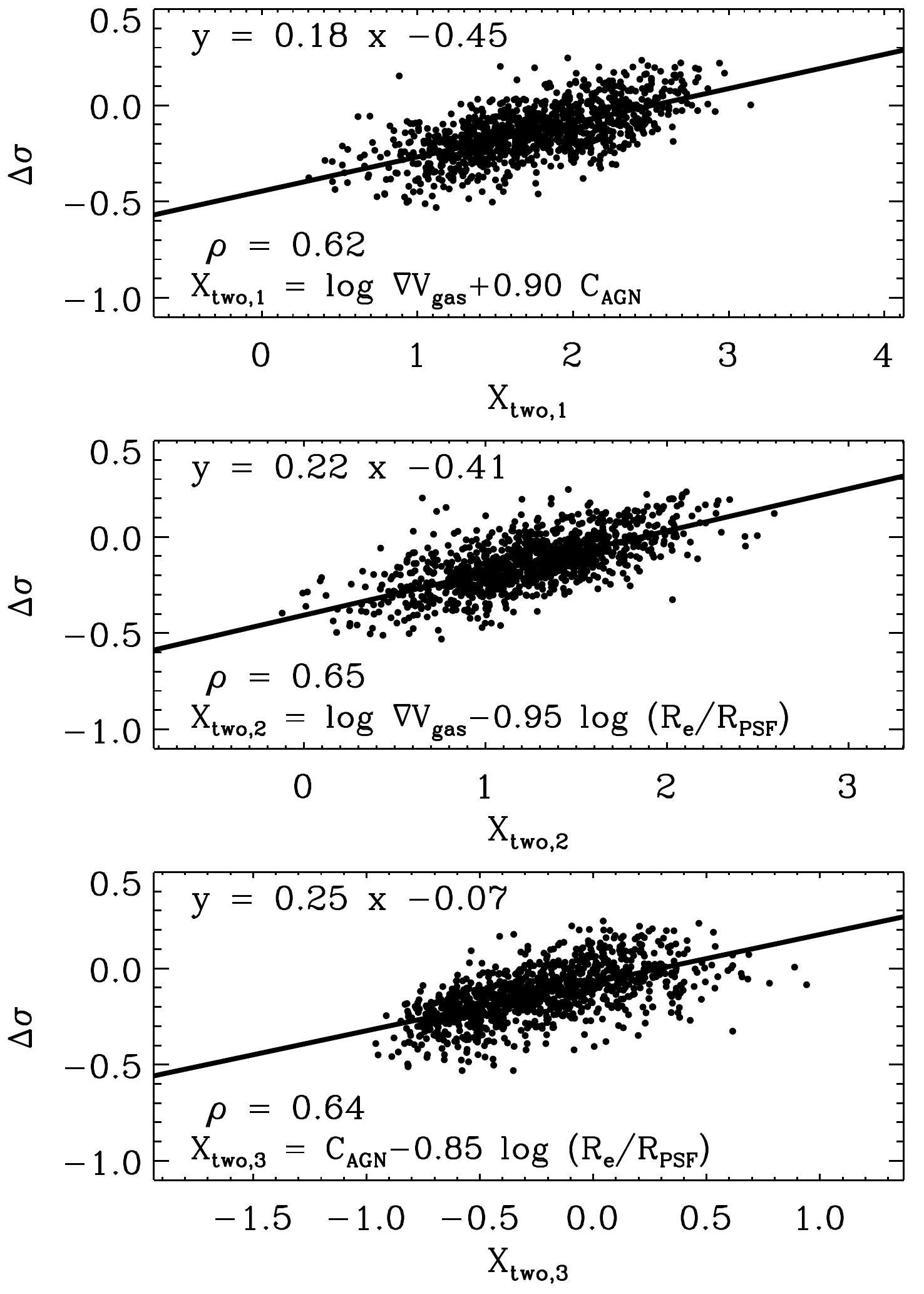}
\caption{Strong positive correlations between $X_{\rm two}$ and \dsig\ ($\rho=0.62$--0.65). We present a linear fit to each relation on the top and Spearman's correlation coefficient $\rho$ on the bottom of each panel. \dxtwo\ is derived from the linear fit to the relation between $X_{\rm two}$ and \dsig.}
\label{xtwo}
\end{figure}          

We found a strong dependence of \dsig\ on $X_{\rm two}$ ($\rho=0.62$--0.65), though it is difficult to measure which combination is more strongly correlated with \dsig\ based on such similar $\rho$ (Figure~\ref{xtwo}). We calculated \dxtwo\ from the linear fit to the relation between $X_{\rm two}$ and \dsig\ as
\begin{align}\label{eq:xtwo}
\Delta\sigma_{X_{\rm two,1}} &= 0.18 X_{\rm two,1} - 0.45~,\nonumber\\
\Delta\sigma_{X_{\rm two,2}} &= 0.22 X_{\rm two,2} - 0.41~,\\
\Delta\sigma_{X_{\rm two,3}} &= 0.25 X_{\rm two,3} - 0.07~.\nonumber
\end{align}

In Table~\ref{tab:xone}, we present $\rho$ for the relations between (\dsig$-$\dxtwo) and the galaxy parameters. Although the correction for $X_{\rm two,1}$, the combination of \rvg\ and \bpt, removes most of the other correlations, we still find a residual correlation between (\dsig$-$$\nabla\sigma_{X_{\rm two,1}}$) and \rre\ ($\rho=-0.37$). $X_{\rm two,2}$, the combination of \rre\ and \rvg, does not fully explain the dependence of \dsig\ on \bpt\ ($\rho=0.22$). A residual correlation ($\rho=0.29$) between (\dsig$-$$\nabla\sigma_{X_{\rm two,3}}$) and log\,\rvg\ is also detected when considering \rre\ and \bpt. In conclusion, the two-parameter corrections still do not fully explain the dependence of \dsig\ on the key galaxy parameters, which implies the three key parameters (\rre, \rvg, and \bpt) independently contribute to the difference between gas and stellar velocity dispersions. 

\subsubsection{Three primary parameters}
\label{sec:xthree}

\begin{figure}
%\centering\includegraphics[width=\columnwidth]{fig8.eps}
\centering\includegraphics[width=\columnwidth]{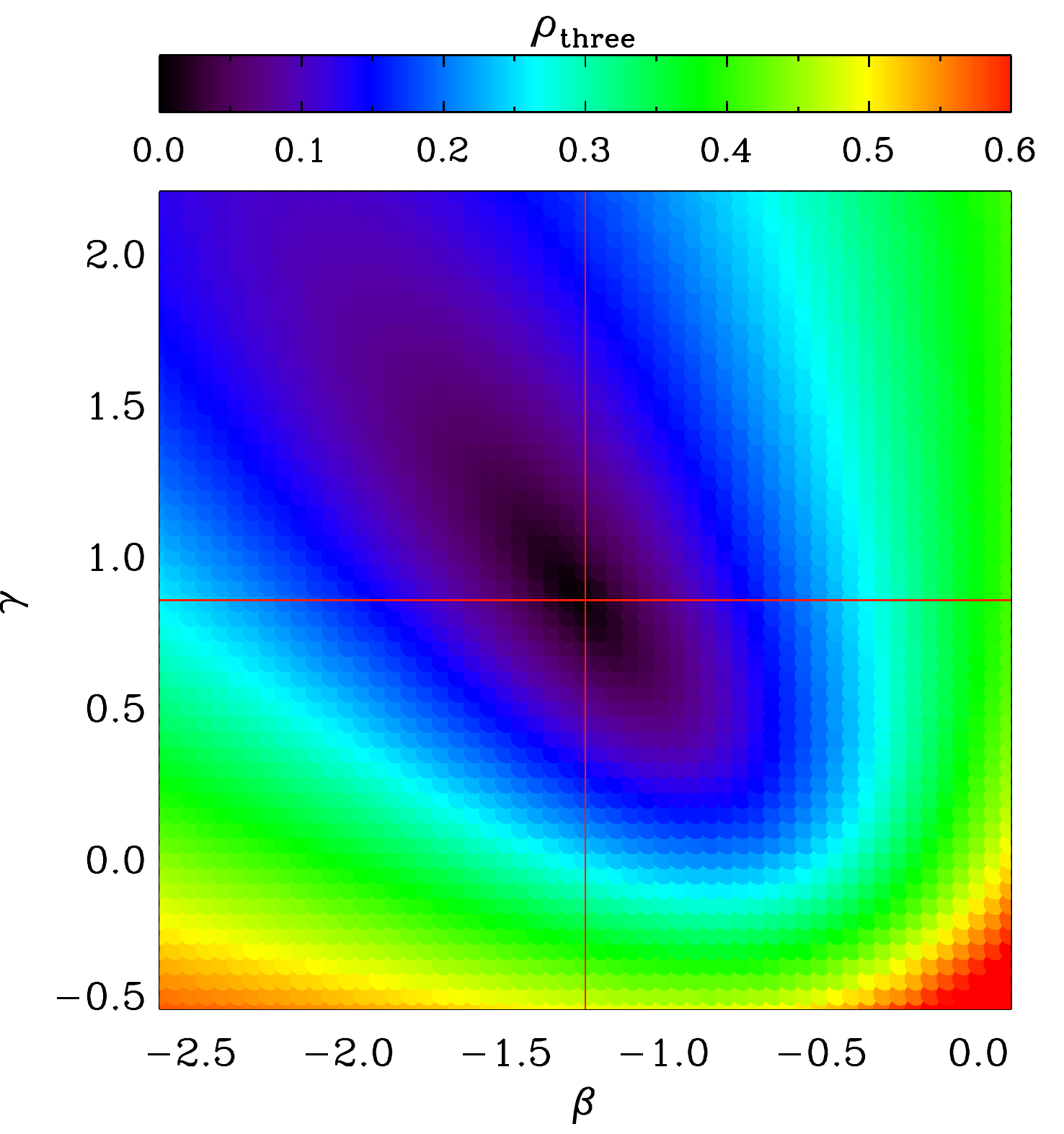}
\caption{Optimal $\beta$ and $\gamma$ for $X_{\rm three} = \log$\rvg$\,+\,\beta \log$(\rre)$\,+\,\gamma$\bpt\ are obtained by minimising $\rho_{\rm three} = \sqrt{\rho_1^2 + \rho_2^2 + \rho_3^2}$, where $\rho_1$, $\rho_2$ and $\rho_3$ are, respectively, calculated from the log\,\rvg--(\dsig$-$$\Delta\sigma_{X_{\rm three}}$), log(\rre)--(\dsig$-$$\Delta\sigma_{X_{\rm three}}$), and \bpt--(\dsig$-$$\Delta\sigma_{X_{\rm three}}$) relations. $\Delta\sigma_{X_{\rm three}}$ is measured using a linear fit to the relation between \dsig\ and $X_{\rm three}$ (see Figure~\ref{xthree}).}
\label{rho_xthree}
\end{figure}

\begin{figure}
%\centering\includegraphics[width=\columnwidth]{fig9.eps}
\centering\includegraphics[width=\columnwidth]{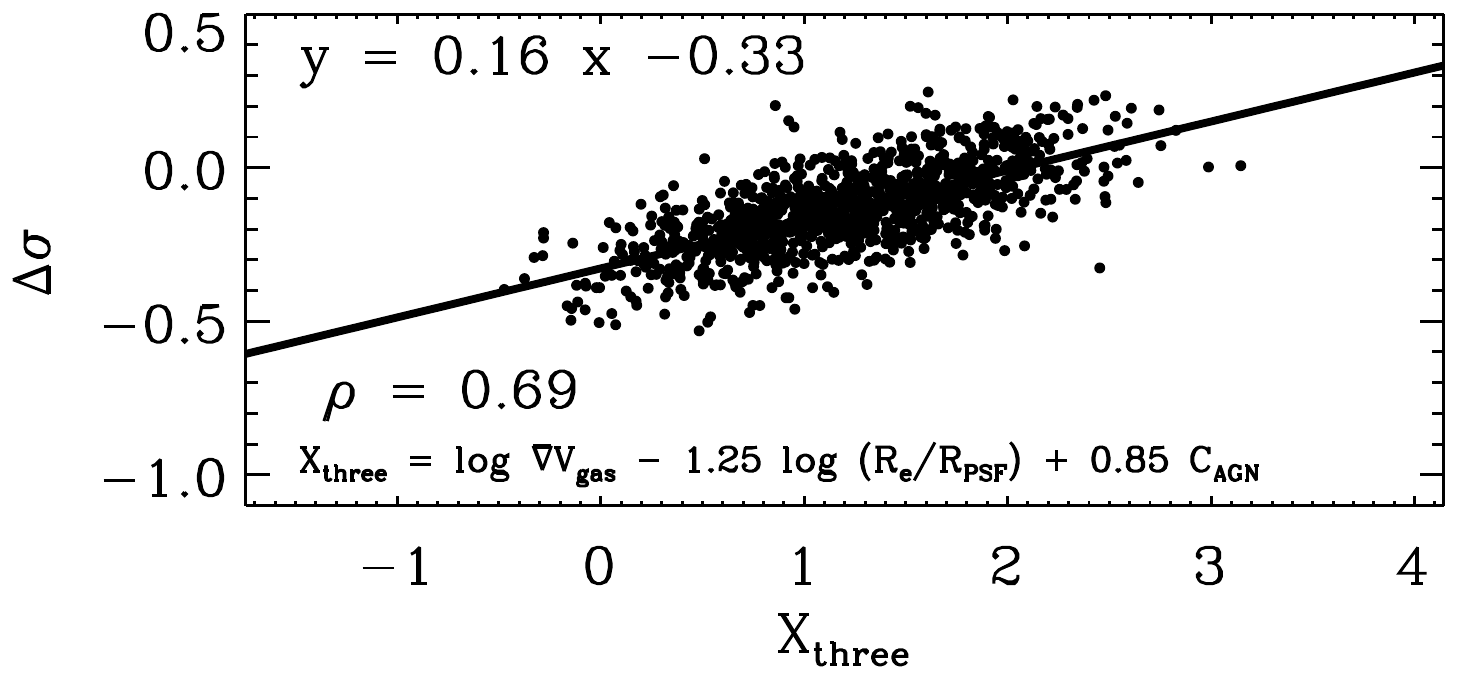}
\caption{The relation that best explains the correlation between \dsig\ and the quantity $X_{\rm three}$, defined as the linear combination of galaxy parameters $\log$\rvg$\,-\,1.25\log$(\rre)$\,+\,0.85$\bpt, is $\Delta\sigma_{X_{\rm three}} = 0.16 X_{\rm three} - 0.33$.}
\label{xthree}
\end{figure}

We examined a third approach assuming all of \rre, \rvg, and \bpt\ are at least partially responsible for the difference in the gas and stellar velocity dispersions. We applied the same logic as above and defined $X_{\rm three} = \log$\rvg$\,+\,\beta \log$(\rre)$\,+\,\gamma$\bpt, a linear combination of the three parameters with contribution factors $\beta$ and $\gamma$. We then numerically estimated the $\beta$ and $\gamma$ that minimise a quadrature sum of Spearman correlation coefficients, $\rho_{\rm three} = \sqrt{\rho_1^2 + \rho_2^2 + \rho_3^2}$ where $\rho_1$, $\rho_2$, and $\rho_3$ are, respectively, calculated from the log\,\rvg--(\dsig$-$$\Delta\sigma_{\rm Xthree}$), log\,(\rre)--(\dsig$-$$\Delta\sigma_{\rm Xthree}$), and \bpt--(\dsig$-$$\Delta\sigma_{\rm Xthree}$) relations. We present the change in $\rho_{\rm three}$ when varying $\beta$ and $\gamma$ in Figure~\ref{rho_xthree}. The contribution factors $\beta$ and $\gamma$ were set using the global minimum of $\rho_{\rm three}$. Finally, we found $X_{\rm three} = \log$\rvg$\,-\,1.25\log$(\rre)$\,+\,0.85$\bpt, which best explains the correlations between \dsig\ and galaxy parameters. We accordingly found a strong correlation ($\rho = 0.69$) between $X_{\rm three}$ and \dsig\ (Figure~\ref{xthree}). 

\begin{figure*}
%\centering\includegraphics[width=\textwidth]{fig10.eps}
\centering\includegraphics[width=\textwidth]{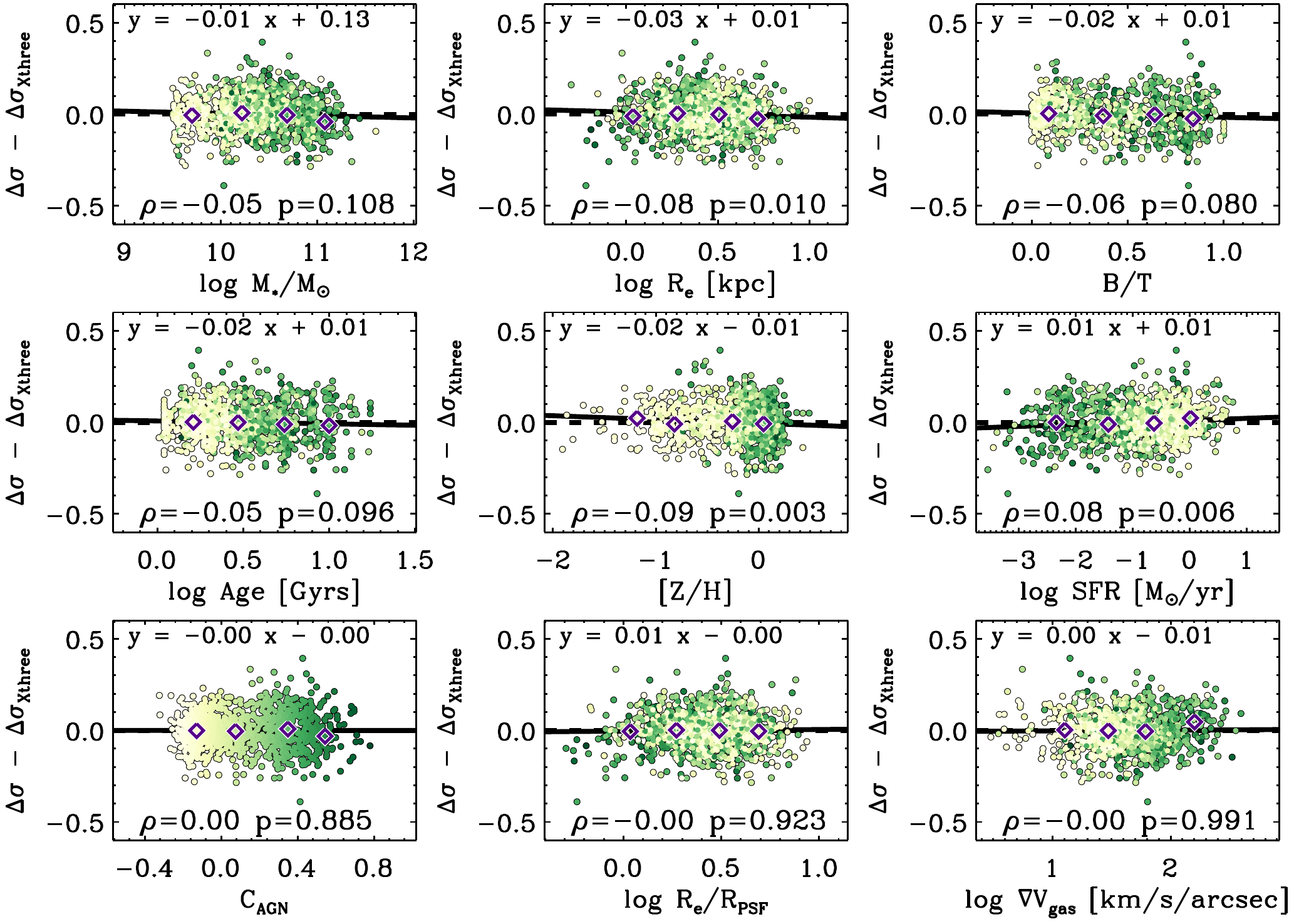}
\caption{The relations between \dsig$-$$\Delta\sigma_{\rm Xthree}$ and the galaxy parameters. The correction using $X_{\rm three}$, combining \rvg, \rre, and \bpt, effectively removes the dependence of \dsig\ on the galaxy parameters. Details are the same as described in Figure~\ref{sigdiff}.}
\label{sigthree}
\end{figure*}

In Figure~\ref{sigthree}, we present the residual correlation after applying the correction $\Delta\sigma_{X_{\rm three}} = 0.16 X_{\rm three} - 0.33$. The low values of the Spearman correlation coefficients ($|\rho| \lesssim 0.1$) confirm that the correction using the three key parameters effectively removes the dependence of \dsig\ on the galaxy parameters. We also found the correction using $\Delta\sigma_{X_{\rm three}}$ removes the dependence of \dsig\ on other galaxy parameters (e.g., $g-i$ colour, log($M_\stars$/\re), log($M_\stars$/\re$^2$), ellipticity, inclination, and S\'ersic $n$), although for brevity we do not present it. We also explored various combinations of galaxy parameters for making $X_{\rm two}$ and $X_{\rm three}$, but none produced a comparable result with $X_{\rm three} = \log$\rvg$\,-\,1.25\log$(\rre)$\,+\,0.85$\bpt. The other combinations for $X_{\rm two}$ and $X_{\rm three}$ yield $\rho$ greater than at least 0.23 for the residual correlations between the corrected \dsig\ and the galaxy parameters. 

The analysis in this and previous sections suggest that all of \rre, \rvg, and \bpt\ are required to fully explain the dependence of \dsig\ on galaxy properties. The result is also supported by partial least squares (PLS) regression method shown in Appendix~\ref{sec:pls}. However this conclusion has been derived using a sample that is biased toward galaxies showing strong emission (Figure~\ref{sam}). We therefore tested the impact of the sampling bias by weighting each galaxy with the reciprocal of the sampling completeness shown in Figure~\ref{sam}. The difference between weighted and unweighted Spearman correlation coefficients was only 0.01; we thus expect marginal impact from sampling bias on the results. Another caveat to this analysis is that we assumed a linear relationship between \dsig\ and the galaxy parameters, which may introduce bias in the result. 

\section{Discussion}

\subsection{Beam smearing}
\label{sec:beam}
As mentioned in Section~\ref{sec:xone}, \dsig\ correlates more strongly with \rre\ than \re, which implies that \dsig\ has a closer relationship to the observational limitation (i.e.\ beam smearing) than the physical size of galaxies. Beam smearing spatially smooths kinematics from IFS data, reducing the velocity gradient and increasing the velocity dispersion. The magnitude of beam smearing depends on both $R_{\rm PSF}$ and $\nabla V$\footnote{In principle, the magnitude of beam smearing also depends on the steepness of the light profile. However, the correction using $\Delta\sigma_{\rm Xthree}$ removes the correlation between \dsig\ and S\'ersic $n$ (with $\rho=0.28$).}. That is, large $R_{\rm PSF}$ and/or high $\nabla V$ can boost the impact of beam smearing, artificially inflating the measured velocity dispersions and deflating the rotation velocities. Considering both $\sigma_\stars$ and $\sigma_\gas$ have been averaged within \re, and $R_{\rm PSF}$ is 1--1.5$^{\prime\prime}$ for typical SAMI observations, the magnitude of beam smearing on $\sigma_\stars$ and $\sigma_\gas$ has links to \rre. 

A gas disc supported by rotation is thinner and rotates faster than a stellar disc at the same location (e.g.\ Shetty et~al.\ 2020), making gas kinematics more susceptible to beam smearing. As a result, $\sigma_\gas$ correlates more strongly with both \rre\ and $\log\nabla V$ than $\sigma_\stars$ does; for $\sigma_\gas$, $\rho = 0.82$ and $-0.30$ for $\log\nabla V_\gas$ and log(\rre) respectively, whereas for $\sigma_\stars$, $\rho = 0.43$ and $-0.14$ for $\log\nabla V_\stars$ and log(\rre)  respectively. These results support our interpretation that the dependence of $\Delta\sigma$ on \rre\ and \rvg\ is largely due to the differential impact of beam smearing on $\sigma_\gas$ and $\sigma_\stars$. It is also supported by a strong correlation between \dsig\ and $\log(\nabla V_{\rm gas}/\nabla V_{*})$ (Appendix~\ref{sec:app1}).

In Section~\ref{sec:res}, we showed that \dsig\ is most closely related to \rvg. So far, we have interpreted the velocity gradient as a tracer for the impact of beam smearing. It is obvious that a higher $\nabla V$ is an important factor indicating large beam smearing, but $\nabla V$ is not completely determined due to observational limitations: unlike $R_{\rm PSF}$, $\nabla V$ also reflects the kinematic structure of galaxies. We presume that there might be underlying reasons for galaxies displaying diverse \rvg, such as gas outflows. However, it is difficult to examine whether additional physical mechanisms are required to explain the dependence of \dsig\ on \rvg~without removing the effect of beam smearing from $\nabla V$. Even if there is a physical reason for the correlation between \dsig\ and \rvg, the correlation becomes artificially stronger due to beam smearing.

The simplest method to rule out the impact of beam smearing is excluding galaxies that are small relative to the size of the PSF. We tested our results excluding the cases with \rre$ < 2$. This weakens the dependence of \dsig\ on \rre, as expected, but there is still a tight correlation between \dsig\ and \rvg. Also, we still find a correlation between \dsig\ and \bpt\ when excluding small galaxies. We measured \dsig\ at 0.9<$R/$\re<1.1, excluding central spaxels where the impact of beam smearing is generally high, and found \dsig\ depends on \rre, \rvg, and \bpt, implying that the kinematics measured at \re\ is still affected by beam smearing.

\begin{figure*} \centering 
\centering\includegraphics[width=\textwidth]{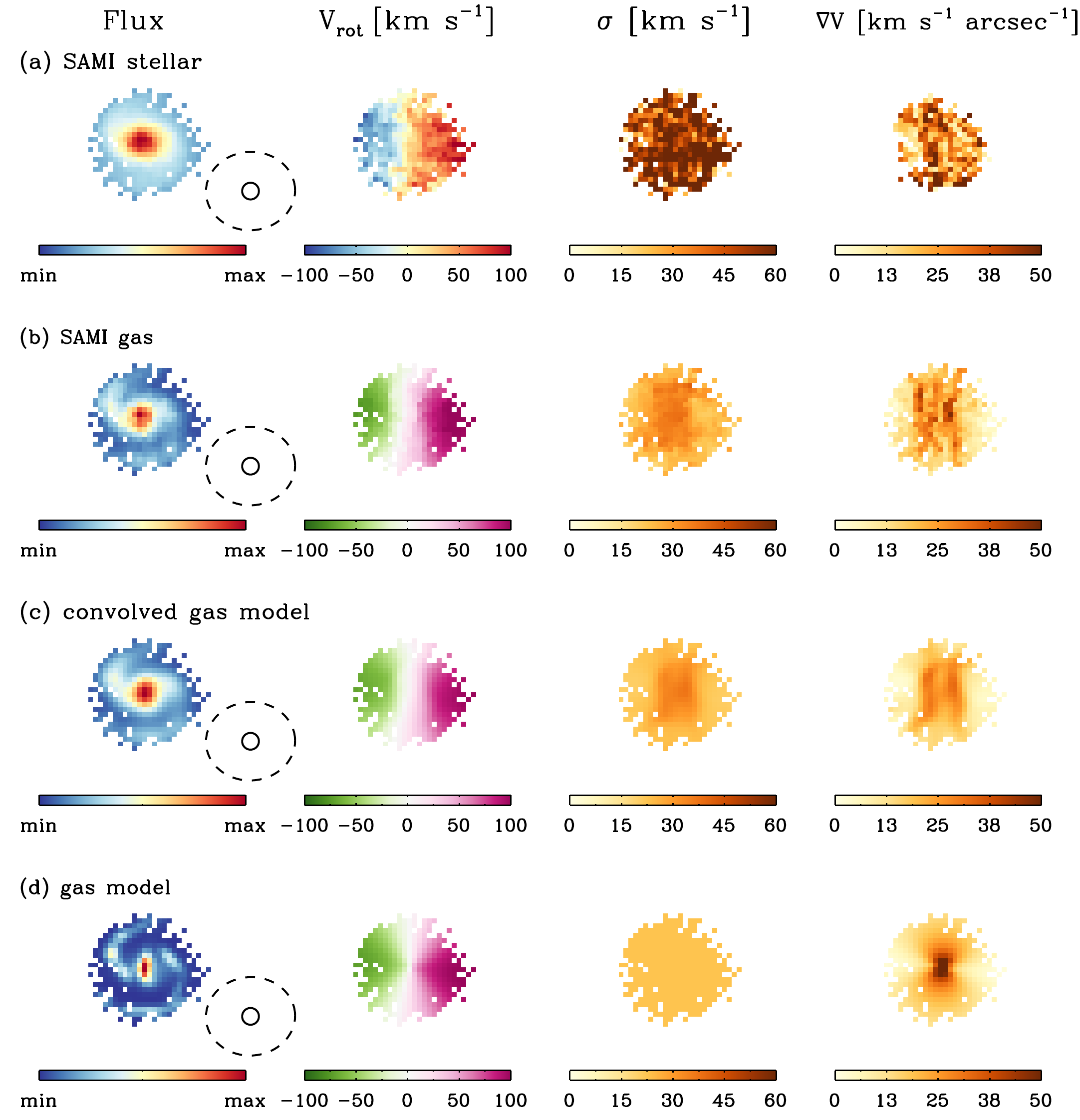}

\caption{(a)~Stellar kinematics and (b)~gas kinematics for a large star-forming galaxy (SAMI ID 485885; \rre\ $=5.33$ and \bpt\ $=-0.21$), illustrating the impact of beam smearing. The ellipse and circle indicate, respectively, \re\ and PSF. The first column is the flux in the continuum (for stars) or H$\alpha$ (for gas). The second, third, and fourth columns show, respectively, the rotation velocity, velocity dispersion, and velocity gradient. The distribution of gas velocity dispersions resembles that of the gas velocity gradients, a sign of beam smearing. (c)~Mock gas kinematics for the galaxy using a hierarchical Gaussian mixture model with PSF convolution. (d)~Gas kinematics for the model before PSF convolution, obtained from Figure~9 of Varidel et~al.\ (2019). Comparing rows (c) and (d), it is apparent that the PSF convolution (i.e.\ beam smearing) artificially inflates the velocity dispersion and produces a correlation between the velocity gradient and the velocity dispersion. Note that Varidel et~al.\ (2019) constructed and tested this modelling approach for star-forming galaxies.}\label{beam}
\end{figure*}

We found other evidence of beam smearing, especially on gas velocity dispersions, even for galaxies that are large compared to the PSF size. In Figure~\ref{beam}, we present gas and stellar kinematics of a large star-forming galaxy with \rre\ $=5.33$ and \bpt\ $=-0.21$, showing that the distribution of the gas velocity dispersion resembles that of the gas velocity gradient. We present mock gas kinematics for the galaxy using a hierarchical Gaussian mixture model obtained from Figure~9 of Varidel et~al.\ (2019). The beam smearing artificially inflates the velocity dispersion, highlighting a resemblance between the velocity gradient and the velocity dispersion for the sample galaxy (Figure~\ref{beam}c). We conclude that limiting galaxy size only partially eliminates the impact of beam smearing, especially on gas velocity dispersions; we still need to consider beam smearing even for large galaxies if they are expected to show a steep velocity gradient (e.g.\ highly-inclined galaxies or compact H$\alpha$ emission).

As introduced in Figure~\ref{beam}, several approaches have been proposed for a beam smearing correction or a full disc modelling to recover intrinsic kinematics (Green et~al.\ 2010; Bouch$\acute{\rm e}$ et~al.\ 2015; Di Teodoro \& Fraternali 2015; Bekiaris et~al.\ 2016; Varidel et~al.\ 2016, 2019; Federrath et~al.\ 2017; Johnson et~al.\ 2018). Most of these methods have focussed on accounting for gas kinematics from star-forming disc-dominant samples (but see Harborne et~al.\ 2020). It has not been confirmed whether the same beam smearing correction (or modelling approach) applies to both gas and stellar kinematics, especially for non-star-forming populations where gas emission is not limited to the star-forming disc. While beam smearing is an essential (but non-physical) factor to be considered when comparing gas and stellar kinematics, current approaches to correct and model the effect of beam smearing still need to be tested for both gas and stellar kinematics from various populations including early-type quiescent galaxies and AGNs. Dynamical modelling (e.g.\ Jeans anisotropic modelling; Cappellari 2008) may be used to correct the stellar kinematics for beam smearing, but lies outside the scope of this work.

\subsection{Gas outflows}

We find that there still is a correlation between \dsig\ and \bpt\ even after accounting for the impact of beam smearing, suggesting a physical mechanism associated with \bpt\ is contributing to the difference between stellar and gas velocity dispersions. This result is consistent with a previous study comparing central gas and stellar velocity dispersions (Ho~2009).

Gas outflows and shocks are effective ways to change \dsig, having a larger impact on the gas kinematics than the stellar kinematics. Ho~et~al.\ (2014) spectrally decomposed emission-line profiles from SAMI IFS data to separate different kinematic components in the line-of-sight direction and suggested shock excitation explains a broad kinematic component. They modelled radiative shocks using a shock/photoionization code, MAPPINGS (Sutherland \& Dopita 1993; Dopita et~al.\ 2013) and estimated the potential contribution and magnitude of shocks for the given optical line ratios. They illustrated the expected shock fraction in the BPT diagram, where a strong positive correlation is readily predictable between the shock fraction and the \bpt\ parameter in this study (Figure~13 in Ho~et~al.\ 2014; see also D'Agostino et~al.\ 2019). 

AGN often generates galaxy-scale outflows (e.g.\ Nesvadba et~al.\ 2006; Sharp \& Bland-Hawthorn 2010; DeBuhr et~al.\ 2012; Cicone et~al.\ 2014; Harrison et~al.\ 2014; Karouzos et~al.\ 2016), which may boost gas velocity dispersions and produce a dependence of \dsig\ on \bpt. Starburst-driven outflows can also affect the structure and kinematics of the ionised gas (see Veilleux, Cecil \& Bland-Hawthorn 2005 for a review). They are found to be common in galaxies with high SFR (Heckman 2002; Rupke, Veilleux \& Sanders 2005a,b; Weiner et~al.\ 2009; Steidel et~al.\ 2010). Avery et~al.\ (2021) also identified evidence of outflows in 322 of 2744 galaxies using MaNGA IFS data, and reported strong correlations between mass outflow rates and both star formation rate and AGN luminosity. In this study, however, we found a negative correlation between \dsig\ and SFR (Figure~\ref{sigdiff}), suggesting the current status of star-forming activity may not be as important in boosting $\sigma_\gas$ and determining \dsig. Another explanation is that star-formation driven outflows are less powerful than AGN driven outflows, generating only a small boost in $\sigma_\gas$; Varidel et~al.\ (2020) found an upturn in $\sigma_\gas$ only for galaxies with very high SFRs. \bpt, which quantifies the contribution of AGN using emission line diagnostics, is also expected to be a proxy for the Eddington ratio, which quantifies the efficiency of black hole accretion and the power of AGN (e.g.\ Jones et~al.\ 2016; Oh~et~al.\ 2017, 2019). We therefore expect to find more frequent and/or massive gas outflows and shocks from galaxies with high values of \bpt. 

\subsection{LINER-like emission and retired galaxies}
\label{sec:liner}
The explanation in the previous section applies when the ionisation sources of the AGN-like emission are indeed supermassive black holes (e.g.\ Seyfert galaxies). However, the majority of local AGN exhibit spectra dominated by emission lines from low-ionisation species ([O\,{\small I}],  [N\,{\small II}],  [S\,{\small II}]), classifying them as low-ionisation nuclear emission-line regions (LINERs; Heckman 1980; Ho, Filippenko \& Sargent 1997a; Cid Fernandes et~al.\ 2010; Singh et~al.\ 2013; Leslie et~al.\ 2016). We examine the fraction of Seyfert and LINER candidates among galaxies with AGN-like emission in this study. The [N\,{\small II}]-based emission-line diagnostics that are used to derive \bpt\ are less sensitive in separating Seyfert-like and LINER-like emission, so both Seyfert and LINER candidates have similar \bpt. We identify Seyfert-like and LINER-like emission using the demarcation proposed by Kewley et~al.\ (2006) based on [S\,{\small II}]/H$\alpha$ and [O\,{\small III}]/H$\beta$ line ratios. We find that 35\% of the sample show AGN-like emission and, of these, 85\% are LINER candidates according to the [S\,{\small II}]-based diagnostics. 

Several mechanisms have been proposed for LINER-like emission. LINERs have classically been thought to be associated with low-luminosity AGNs since Heckman (1980) found their emission-line ratios cannot be achieved via star formation alone. Kewley et al. (2006) suggests LINER-like emission can be produced by inefficiently accreting AGN. Some LINER candidates exhibit a central point-like source in X-rays, supporting the suggestion they are powered by AGN (Gonzalez-Martin et~al.\ 2009; M\'{a}rquez et~al.\ 2017). However, the hypothesis regarding LINER-like emission as a subclass of the AGN family has been challenged by finding extended LINER-like emission (Sarzi et~al.\ 2010; Yan \& Blanton 2012; Singh et~al.\ 2013) and overall low H$\alpha$ equivalent widths (EW(H$\alpha$); Cid Fernandes et~al.\ 2010). It turns out that LINER-like emission can also be achieved through excitation by hot old post-AGB (post-asymptotic giant branch) stars (Singh et~al.\ 2013), shocks/outflows (Dopita \& Sutherland 1996; Dopita et~al.\ 1996; Rich, Kewley \& Dopita 2011, 2014; Ho et~al.\ 2016), or extraplanar diffuse ionised gas often detected in edge-on discs (Hoopes \& Walterbos 2003; Madsen, Reynolds \& Haffner 2006; Voges \& Walterbos 2006; Bregman et~al.\ 2013; Belfiore et~al.\ 2015, 2016; Law et~al.\ 2021).

\begin{figure}
%\centering\includegraphics[width=\columnwidth]{fig13.eps}
\centering\includegraphics[width=\columnwidth]{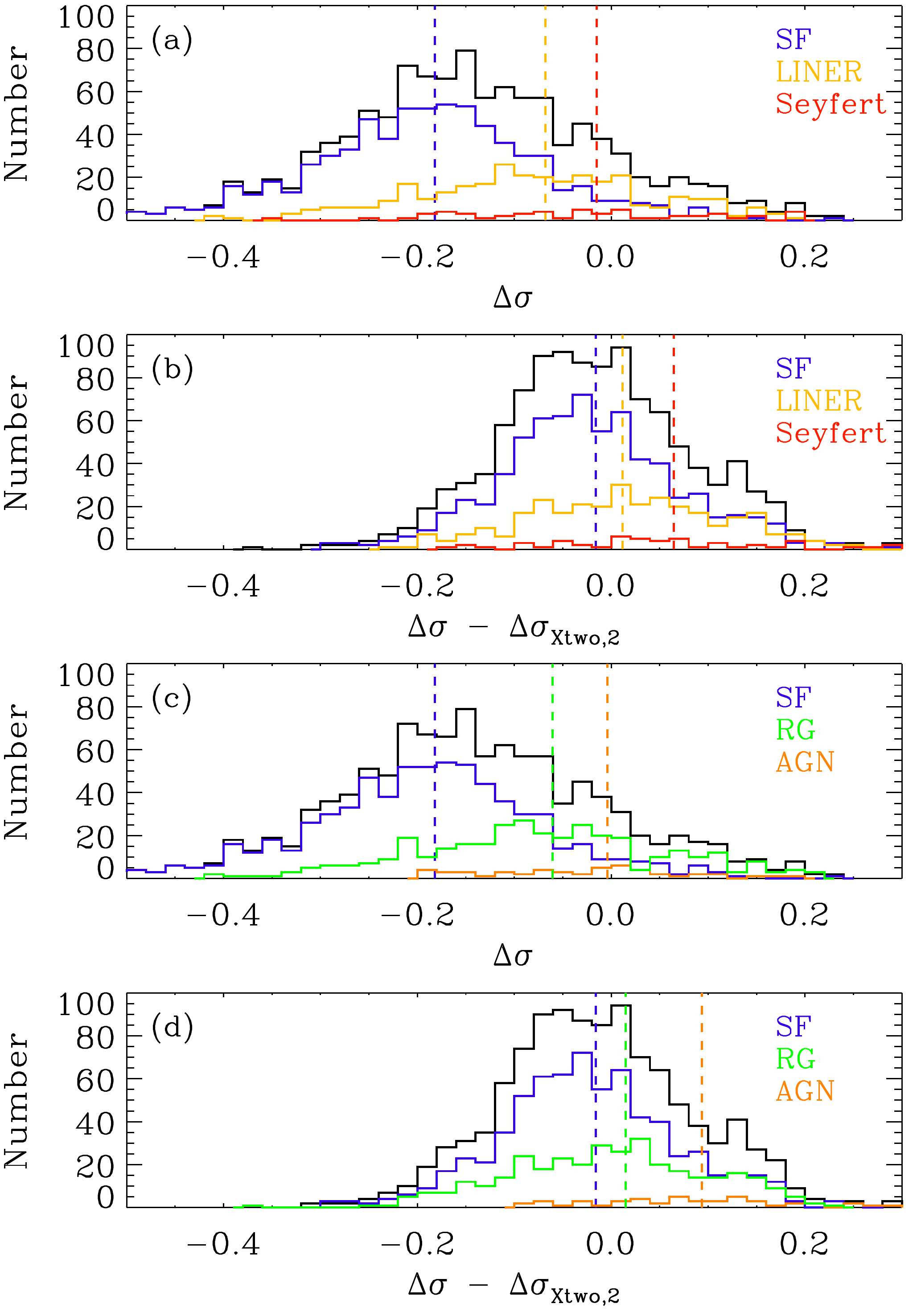}
\caption{(a)~The \dsig\ distribution of emission-line groups, with galaxies classified into star-forming (SF; blue), Seyfert (red), and LINER (yellow) candidates according to the [S\,{\small II}]-based diagnostics (Kewley et~al.\ 2006). (b)~The distribution of $\Delta\sigma - \Delta\sigma_{\rm Xtwo,2}$ for the same groups considering the impact of beam smearing on \dsig. (c)~The \dsig\ distribution of star-forming (SF; blue), AGN (orange), and RG (green) candidates classified using the [S\,{\small II}]-based diagnostics together with EW(H$\alpha$). (d)~The $\Delta\sigma - \Delta\sigma_{\rm Xtwo,2}$ distribution of SF, AGN, and RG groups. LINER (or RG) candidates show \dsig\ and $\Delta\sigma - \Delta\sigma_{\rm Xtwo,2}$ intermediate between SF and Seyfert (or AGN) candidates.}
\label{liner}
\end{figure}

Given the various channels for LINER-like emission, they may also show a wide range of kinematic properties. In Figure~\ref{liner}(a), we present the distribution of \dsig\ for star-forming galaxies (SF), LINERs, and Seyferts classified using the [S\,{\small II}]-based diagnostics (Kewley et~al.\ 2006). Seyfert candidates show higher \dsig\ compared to star-forming galaxies, as expected from the correlation between \dsig\ and \bpt. LINER candidates have slightly lower median \dsig\ than Seyfert candidates, which is a result consistent with the expectation that LINER-like emission may be generated by multiple channels displaying \dsig\ intermediate between star-forming and Seyfert candidates. 

However, as described in Section~\ref{sec:beam}, the discussion on LINER-like emission and \dsig\ is insufficient without considering the impact of beam smearing. We use $\Delta\sigma_{\rm Xtwo,2}$, a correction using the quantity $X_{\rm two,2}$ combining \rvg\ and \rre\ (Equations~5 and~6), to approximately remove the dependence of \dsig\ on beam smearing. We note, however, that it is difficult to measure the precise impact of beam smearing on \dsig\ without properly modelling it; $\Delta\sigma_{\rm Xtwo,2}$ may overestimate the effect of beam smearing on \dsig, considering that \rvg\ is not completely determined due to observational limitations as discussed in Section~\ref{sec:beam}. Despite a conservative correction for beam smearing, LINER candidates display an intermediate $\Delta\sigma - \Delta\sigma_{\rm Xtwo,2}$, between those of star-forming and Seyfert candidates (Figure~\ref{liner}b). A similar result is shown in Figure 4 of Law et al. (2021).

We also explore diagnostics using EW(H$\alpha$) to differentiate retired galaxies (RG) that have stopped forming stars (Cid Fernandes et~al.\ 2010, 2011). Galaxies with EW(H$\alpha$)\,<\,3\AA\ are identified as RG, and galaxies with EW(H$\alpha$)\,>\,3\AA\ are separated into star-forming and AGN candidates using the [S\,{\small II}]-based diagnostics following Kewley et~al.\ (2006). We present (Figure~\ref{liner}c and~d) the distributions of \dsig\ and \dsig$-\Delta\sigma_{\rm Xtwo,2}$ for star-forming, RG, and AGN candidates. RG display \dsig\ and \dsig$-\Delta\sigma_{\rm Xtwo,2}$ intermediate between star-forming and AGN candidates, similar to LINER candidates. The difference between RG and AGN in \dsig$-$$\Delta\sigma_{\rm Xtwo,2}$ is more prominent than the difference between LINER and Seyfert candidates (Figure~\ref{liner}b and~d). 

The link between AGN and $\Delta\sigma$ is not seriously challenged by the fact that the majority of AGN-like emission are identified as LINER (or RG) candidates because Seyfert galaxies tends to show high values of $\Delta\sigma$ compared to any other ionisation sources. However, the result may not rule out an alternative explanation for the enhanced $\Delta\sigma$ in LINER (or RG) considering that the majority of LINER candidates are expected to be powered by hot evolved stars (Belfiore et al. 2016). Law et al. (2021) suggests that a large vertical scale height of diffuse ionised gas powered by evolved stars explains enhanced gas velocity dispersions. In that sense, AGN and evolved stars may work together to enhance gas velocity dispersions, generating a strong correlation between $\Delta\sigma$ and C$_{\rm AGN}$.

\subsection{Emission line regions}
\label{region}

Another possibility explaining the correlation between \dsig\ and \bpt\ is the connection between the location of the ionised gas and the power source: $\sigma_\gas$ may reflect the overall kinematics of the region containing the ionised gas, which changes according to the power sources. The H{\small II} regions in actively star-forming galaxies are concentrated along the (thin) disc structure. This is kinematically colder than the overall stellar kinematics, which can also comprise kinematically hot components (i.e.\ thick disc and bulge structures). Therefore, it is reasonable to find lower $\sigma_\gas$ than $\sigma_\stars$ in star-forming galaxies because $\sigma_\gas$ represents the kinematics of the star-forming region, which is dynamically colder than the overall stellar structure. On the other hand, we may find kinematically warmer diffuse ionised gas from non-star-forming populations (LINER, RG, and possibly AGN) due to the relative weakness of star formation and potential absence of the thin disc. In active galaxies, we find narrow-line regions powered by a supermassive black hole within a few kiloparsecs of the galactic centre, where we expect dynamically hot kinematics.

We tested the concentration of H$\alpha$ emission and stellar continuum of star-forming galaxies, LINER, and Seyfert candidates. The concentration of H$\alpha$ flux (C$_{\rm H\alpha}$) and continuum (C$_{*}$) is estimated as
\begin{equation} C = \frac{F_{0.5Re}}{F_{Re}}~,
\end{equation}
where F$_{\rm 0.5Re}$ and F$_{\rm Re}$ are, respectively, the H$\alpha$ or continuum flux measured at 0.5 and 1\,$R_{\rm e}$, which are derived in $r$-band. Seyfert candidates identified by the [S\,{\small II}]-based diagnostics display a higher median C$_{\rm H\alpha}$ than star-forming galaxies (0.418 versus 0.327); we found the similar median C$_{\rm H\alpha}$ between Seyfert and LINER candidates (0.424). We find the median C$_{*}$ = 0.364, 0.398, and 0.344 for respectively Seyfert, LINER, and star-forming galaxies. Seyfert galaxies present more centrally-concentrated H$\alpha$ emission than their stellar continuum (median(C$_{\rm H\alpha}$/C$_{*}$) = 1.141). LINER candidates also show slightly higher C$_{\rm H\alpha}$ than C$_{*}$ (median(C$_{\rm H\alpha}$/C$_{*}$) = 1.075). On the other hand, the stellar continuum is slightly more centrally concentrated than H$\alpha$ emission in star-forming galaxies (median(C$_{\rm H\alpha}$/C$_{*}$) = 0.965). The results suggest that the dependence of $\Delta\sigma$ on C$_{\rm AGN}$ can also be described by the association between the location of ionised regions and the power source. The diagnostics based on EW(H$\alpha$) yield consistent results.

\subsection{Aperture velocity dispersions}

\begin{table*}\centering\caption{Spearman's correlation coefficient $\rho$ calculated from the relation between \dsig$_\aper$$-$$\Delta\sigma_{\rm X}$ and galaxy parameters}
\begin{tabular}{lrrrrrrrrr}
\hline\hline
\dsig$_{\rm X}$ & log($M_\stars$/$M_{\odot}$) & log\,$R_{\rm e}$ & B/T & log\,Age & $[Z/H]$ & log\,SFR & \bpt & log(\rre) & log\,\rvg \\
\hline
no correction  & 0.23 & -0.06 & 0.13 & 0.28 & 0.24 & -0.16 & 0.32 & -0.11 & 0.51\\
\hline
\dsig$_{\rm log (M_\stars/M_{\odot})}$ & \textbf{0.00} & -0.19 & \textbf{0.08} & 0.19 & \textbf{0.08} & -0.18 & 0.19 & -0.16 & 0.36 \\
\dsig$_{\rm log R_{\rm e}} $ & 0.26 & \textbf{0.01} & 0.12 & 0.27 & 0.25 & -0.13 & 0.32 & \textbf{-0.07} & 0.49 \\
\dsig$_{\rm B/T}$ & 0.19 & \textbf{-0.01} & \textbf{-0.01} & 0.20 & 0.20 & \textbf{-0.10} & 0.26 & \textbf{-0.07} & 0.42 \\
\dsig$_{\rm log Age}$  & 0.12 & \textbf{-0.03} & \textbf{0.03} & \textbf{0.01} & 0.13 & \textbf{0.01} & 0.14 & \textbf{-0.06} & 0.37 \\
\dsig$_{\rm [Z/H]}$  & \textbf{0.04} & -0.11 & \textbf{0.06} & 0.18 & \textbf{-0.02} & \textbf{-0.07} & 0.15 & -0.11 & 0.33 \\
\dsig$_{\rm log SFR}$   & 0.24 & \textbf{0.02} & \textbf{0.08} & 0.19 & 0.19 & \textbf{-0.00} & 0.25 & \textbf{-0.05} & 0.46 \\
\dsig$_{\rm C_{AGN}}$ & \textbf{0.05} & \textbf{-0.04} & \textbf{-0.00} & \textbf{0.07} & \textbf{0.03} & \textbf{0.01} & \textbf{0.02} & \textbf{-0.05} & 0.33 \\
\dsig$_{\rm log R_{\rm e}/R_{\rm PSF}}$  & 0.25 & \textbf{0.03} & \textbf{0.10} & 0.25 & 0.24 & -0.11 & 0.30 & \textbf{0.01} & 0.46 \\
\dsig$_{\rm log \nabla V_\gas}$  & \textbf{-0.10} & \textbf{-0.06} & \textbf{-0.03} & \textbf{0.06} & \textbf{-0.05} & -\textbf{0.07} & \textbf{0.05} & \textbf{-0.00} & \textbf{-0.02} \\
\hline\hline
\multicolumn{10}{l}{Note: the correction factor $\Delta\sigma_{\rm X}$ is derived using a linear fit to the relation between $X$ and \dsig$_\aper$; values in bold if $|\rho| \leq 0.1$.}\\
\label{tab:aper}\end{tabular}
\end{table*}

The connection between \dsig\ and \bpt\ discussed in the previous sections raises the question whether it also applies to the dynamical mass derived from gas and stellar kinematics. We derived $\sigma_\gas$ and $\sigma_\stars$ averaging locally-measured line-of-sight velocity dispersions from spaxels within \re. That is, both $\sigma_\gas$ and $\sigma_\stars$ explicitly exclude {\em rotation} components, and so they are not comparable to the velocity dispersion derived from aperture spectra ($\sigma_\aper$), which is similar to the second-order velocity moment $V_{\rm rms} \equiv \sqrt{\sigma^2 + V_{\rm rot}^2}$ or the kinematic parameter S$_{0.5} \equiv \sqrt{\sigma^2+0.5 V_{\rm rot}^2}$ introduced in Weiner, Willmer \& Faber (2006). Barat et~al.\ (2019) reported an in-depth comparison between the spaxel- and aperture-based velocity dispersions and S$_{0.5}$. The aperture velocity dispersion, combining both rotation and dispersion, is a proxy for the dynamical mass, whereas averages of the local line-of-sight velocity dispersions are not.

We generated aperture spectra combining spectra from the spaxels within an elliptical 1\,\re\ aperture. Then we extracted the gas and stellar velocity dispersions from aperture spectra ($\sigma_{\aper,\gas}$ and $\sigma_{\aper,\stars}$) with {\sc pPXF}, following the same method described in Section~\ref{ppxf}. The difference between the gas and stellar aperture velocity dispersions is defined as \dsig$_\aper \equiv \log \sigma_{\aper,\gas} - \log \sigma_{\aper,\stars}$ in the same manner as Section~\ref{global}. When comparing \dsig\ and \dsig$_\aper$, we detect a significant difference in the median values of \dsig\ and \dsig$_\aper$ ($-0.14$ versus $-0.03$\,dex), suggesting overall gas has smaller local line-of-sight dispersions compared to stars, whereas gas and stars show similar dispersions integrated over an aperture. \dsig$_\aper$ also has a smaller standard deviation (0.10\,dex) than \dsig\ (0.14\,dex). The small differences between $\sigma_{\aper,\gas}$ and $\sigma_{\aper,\stars}$ have also been reported in previous studies (Nelson \& Whittle 1996; Greene \& Ho 2005; Chen, Hao, \& Wang 2008; Gilhuly, Courteau, \& Sa\'nchez et~al.\ 2019), including at large look-back times (Bezanson et~al.\ 2018). 

\begin{figure*}
%\centering\includegraphics[width=\textwidth]{fig14.eps}
\centering\includegraphics[width=\textwidth]{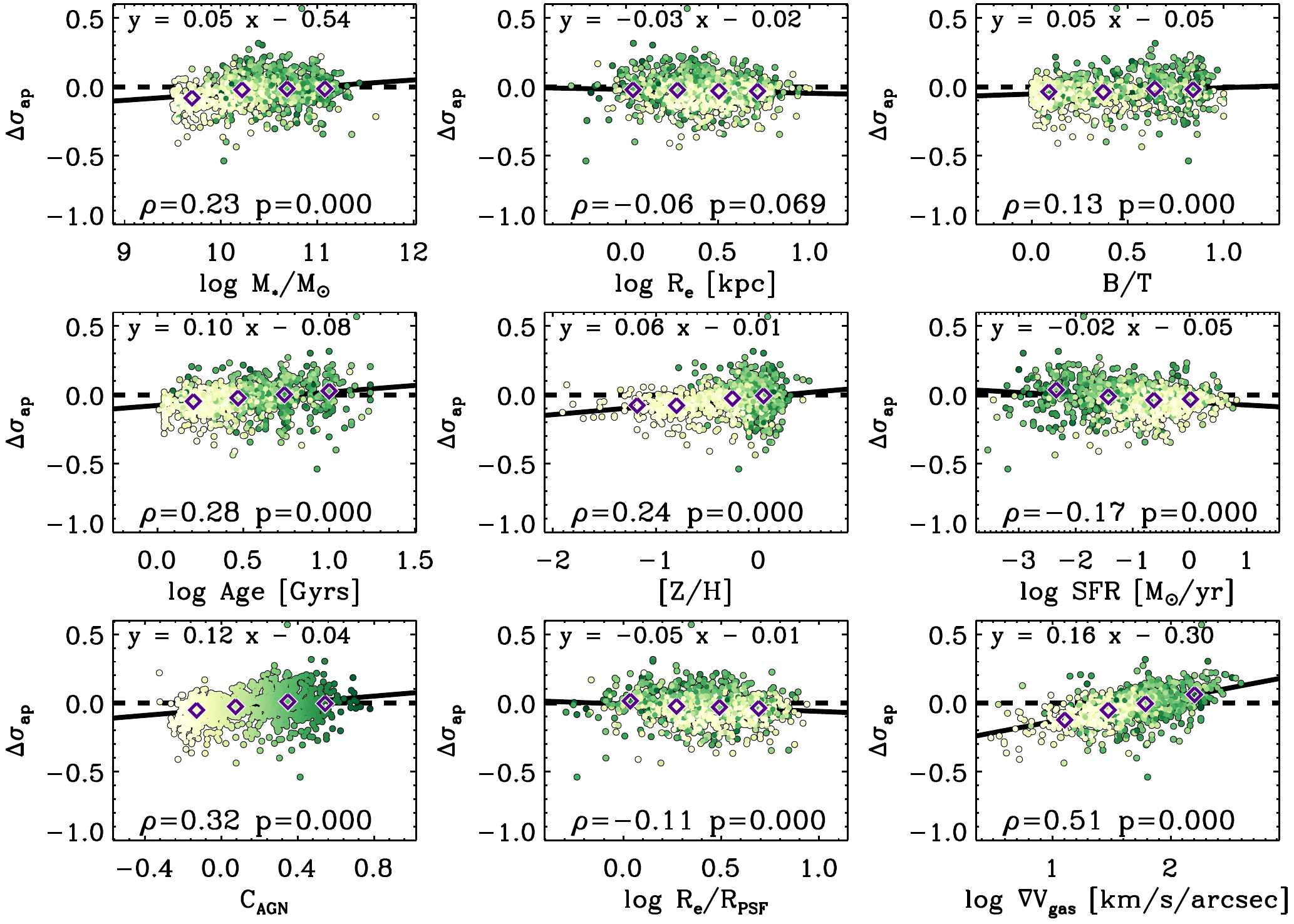}
\caption{The dependence of \dsig$_\aper$ on various galaxy parameters; details are the same as in Figure~\ref{sigdiff}.}
\label{sigaperdiff}
\end{figure*}

We present the Spearman correlation coefficients, $\rho$, for the relations between \dsig$_\aper$ and various galaxy parameters to investigate whether \dsig$_\aper$ in Figure~\ref{sigaperdiff} and Table~\ref{tab:aper}. The absolute values of $\rho$ from the \dsig$_\aper$ relations are smaller than those from the \dsig\ relations (cf.\ Table~\ref{tab:xone}), suggesting \dsig$_\aper$ is less dependent on galaxy parameters than \dsig. \dsig\ shows a strong correlation with \bpt\ ($\rho=0.54$), whereas we find a less prominent dependence of \dsig$_\aper$ on \bpt\ ($\rho = 0.32$). We expect to find little dependence of \dsig$_\aper$ on beam smearing, considering rotations and dispersions are thoroughly combined in \dsig$_\aper$ measured within \re; and in fact \dsig$_\aper$ does not correlate with \rre\ ($\rho = -0.11$). However, we still find a strong correlation between \dsig$_\aper$ and \rvg\ ($\rho = 0.51$), suggesting \rvg\ is not determined solely by beam smearing. We applied the same analysis described in Section~\ref{sec:xone} to \dsig$_\aper$ and found that the correction using \rvg\ yields nearly zero $\rho$ in all the relations, indicating \rvg\ is the only primary parameter having a causal connection to \dsig$_\aper$. The PLS test also finds that \rvg\ is the parameter that explains the most variance in \dsig$_\aper$ (Appendix~\ref{sec:pls}). In summary, \dsig\ depends on \rre, \rvg, and \bpt, but only \rvg\ seem to have a strong connection to \dsig$_\aper$. The physical mechanism underlying \bpt\ does not seem to significantly affect the dynamical mass derived from ionised gas. 

The different behaviour of \dsig\ and \dsig$_\aper$ mainly stems from the difference between $\sigma_\gas$ and $\sigma_{\aper,\gas}$. The medians of $\sigma_{\aper,\gas}/\sigma_\gas$ and $\sigma_{\aper,\stars}/\sigma_\stars$ are, respectively, 1.31 and 1.06. We suspect the rotation component included in $\sigma_{\aper,\gas}$ reduces the dependence of \dsig$_\aper$ on \bpt. $\sigma_\gas$ embeds `local' signatures of gas motions as it is the average of the locally-measured gas velocity dispersion. The local signatures shown in $\sigma_\gas$ may be diminished in $\sigma_{\aper,\gas}$, which integrates dispersions and rotations over the aperture. There is also the counterbalance between rotations and dispersions: galaxies with high \bpt\ (i.e.\ AGN-like emission) tend to be dominated by dispersion, whereas rotation is more prominent in galaxies with low \bpt\ (i.e.\ star-forming galaxies). As a result, star-forming galaxies show a higher $\sigma_{\aper,\gas}/\sigma_\gas$ than galaxies showing AGN-like emission (1.41 versus 1.12).

%The fact that gas and stars show similar aperture velocity dispersions favours the scenario involving different ionised regions according to the power source described in Section~\ref{region}, suggesting dynamically cold (hot) emission from star-forming galaxies (galaxies with AGN-like emission). However, the similar dynamical masses from gas and stars may not rule out the existence of outflows. Ho~et~al.\ (2014) found that the broad-line component tracing outflows is more centrally concentrated and dynamically hotter than the narrow-line (star-forming) component, so that outflows and star-forming components can still indicate the same dynamical potential. 

Although the variation of \dsig$_\aper$ is much smaller than that of \dsig, we find a strong correlation between \dsig$_\aper$ and \rvg, which may introduce correlations between \dsig$_\aper$ and other galaxy parameters (e.g.\ $M_\stars$, Age, $[Z/H]$, and \bpt). There might be an underlying physical effect involved in \rvg\ beyond beam smearing, and one possibility is that steep velocity gradients in the gas trace outflows/shocks (Shih et~al.\ 2013). Although it is unclear what causes the dependence of \dsig$_\aper$ on \rvg, dynamical masses derived from stars and gas may show measurable differences, as shown in Figure~\ref{sigaperdiff}.

\section{Summary and conclusion}

We examined the ionised gas and stellar velocity dispersions for 1090 galaxies from the SAMI galaxy survey, estimated using the emission-line fitting code {\sc lzifu} and the full spectral fitting method {\sc pPXF} (see Section~\ref{method}). The gas and stellar velocity dispersions have been measured as the average of the local velocity dispersions from spatially-resolved spectra within 1\,\re. 

The difference between the (log) velocity dispersions of gas and stars, \dsig, is correlated with various galaxy parameters listed in Section~\ref{sec:par} (Figure~\ref{sigdiff}). This implies that there are one or more physical mechanisms making the ionised gas kinematics different from the stellar kinematics. However, it is not trivial to determine which parameters have a causal connection to \dsig, since galaxy properties often correlate with each other. This leads us to explore two independent methods. We first investigated the combination of multiple parameters that shows the best correlation with \dsig\ while also explaining the relations between \dsig\ and individual galaxy parameters (Section~\ref{sec:comb}). As an alternative, we employed the partial least squares method, which finds the principal parameters explaining the variance in \dsig\ (Section~\ref{sec:pls}). The two methods produced consistent results indicating that \bpt, \rre, and \rvg\ have strong connections with \dsig.

Beam smearing is closely related to both \rre\ and \rvg, and must be considered when discussing gas and stellar kinematics. Excluding small galaxies eliminates many cases and diminishes the correlation between \dsig\ and \rre. However, it does not provide a complete solution to beam smearing, especially for gas kinematics with large velocity gradients. Current approaches to dealing with beam smearing mainly focus on recovering the intrinsic kinematics of extremely thin gas discs. These approaches need to be extended to a broader range of galaxy types (e.g.\ early-type galaxies) and power sources (e.g.\ AGN) for a comprehensive understanding of the impact of beam smearing on gas and stellar kinematics. 
 
The dependence of \dsig\ on \bpt\ suggests that the kinematics of ionised gas are more sensitive to the power sources of gas emission than are the stellar kinematics. $\sigma_\gas$ is much lower than $\sigma_\stars$ in star-forming galaxies, whereas AGN and LINER show $\sigma_\gas$ as comparable to $\sigma_\stars$. Gas outflows powered by AGN may inflate gas velocity dispersions, explaining the correlation between \dsig\ and \bpt. Avery et~al.\ (2021), however, reported that the mass outflow rates depend on the luminosity of AGN and SFR, indicating that both starbursts and AGN are physical drivers of galactic outflows. In this study, we do not find a connection between the current SFR (or SFR normalised by the main sequence, following Avery et~al.\ 2021) and \dsig\ (see Varidel et~al.\ 2020), so there is no direct evidence linking the physical drivers of gas outflows and the impact of outflows on gas kinematics. Another possible explanation for enhanced $\Delta\sigma$, especially valid for LINER, is dynamically-warm diffuse ionised gas powered by evolved stars (Belfiore et al. 2016; Law et al. 2021).

The gas kinematics may reflect the overall motions within the ionised regions. The centrally-concentrated ionised regions of AGN and LINER may produce dynamically-hotter gas kinematics compared to actively star-forming galaxies that show extended H{\small II} regions spreading over the (thin) disc. The compact distribution of H$\alpha$ emission can also lead to a greater impact of beam smearing for AGN and LINER candidates.

The difference between gas and stellar kinematics becomes less prominent when comparing velocity dispersions measured from integrated spectra, which are better proxies for the dynamical mass, integrating both rotation and dispersion components. Thus the mechanism that generates the dependence of \dsig\ on \bpt\ does not seriously violate the dynamical equilibrium of the ionised gas. Nevertheless, there still is a dependence of \dsig$_\aper$ on \rvg, implying a small bias in gas and stellar dynamical masses with \rvg\ that might be related to underlying physics rather than beam smearing. 

\section*{Acknowledgements}

This research was supported by the Australian Research Council Centre of Excellence for All Sky Astrophysics in 3 Dimensions (ASTRO 3D), through project number CE170100013. The SAMI Galaxy Survey is based on observations made at the Anglo-Australian Telescope. The SAMI was developed jointly by the University of Sydney and the Australian Astronomical Observatory. The SAMI input catalogue is based on data taken from the SDSS, the GAMA Survey, and the VST ATLAS Survey. The SAMI Galaxy Survey is supported by the Australian Research Council ASTRO 3D, through project number CE170100013, the Australian Research Council Centre of Excellence for All-sky Astrophysics (CAASTRO), through project number CE110001020, and other participating institutions. The SAMI Galaxy Survey website is sami-survey.org. This research used {\sc pPXF} method by Cappellari \& Emsellem (2004) as upgraded in Cappellari (2017). S.O. thank D.L for the constant support and love. F.D.E. acknowledges funding through the H2020 ERC Consolidator Grant 683184. B.G. is the recipient of an Australian Research Council Future Fellowship (FT140101202). J.v.d.S. acknowledges support of an Australian Research Council Discovery Early Career Research Award (project number DE200100461) funded by the Australian Government. L.C. acknowledges support from the Australian Research Council Discovery Project and Future Fellowship funding schemes (DP210100337, FT180100066). J.J.B. acknowledges support of an Australian Research Council Future Fellowship (FT180100231). S.B. acknowledges funding support from the Australian Research Council through a Future Fellowship (FT140101166). A.M.M. acknowledges support from the National Science Foundation under Grant No. 2009416. M.S.O. acknowledges the funding support from the Australian Research Council through a Future Fellowship (FT140100255). S.K.Y. acknowledges support from the Korean National Research Foundation (NRF-2020R1A2C3003769).

\section*{Data Availability}
The data underlying this article are available from Astronomical Optics' Data Central service at https://datacentral.org.au/ as part of the SAMI Galaxy Survey Data Release 3.

\appendix
\section{\dsig\ and $\nabla V_{\rm \lowercase{gas}}/\nabla V_{*}$.}
\label{sec:app1}
\begin{figure}
\centering
\includegraphics[width=\columnwidth]{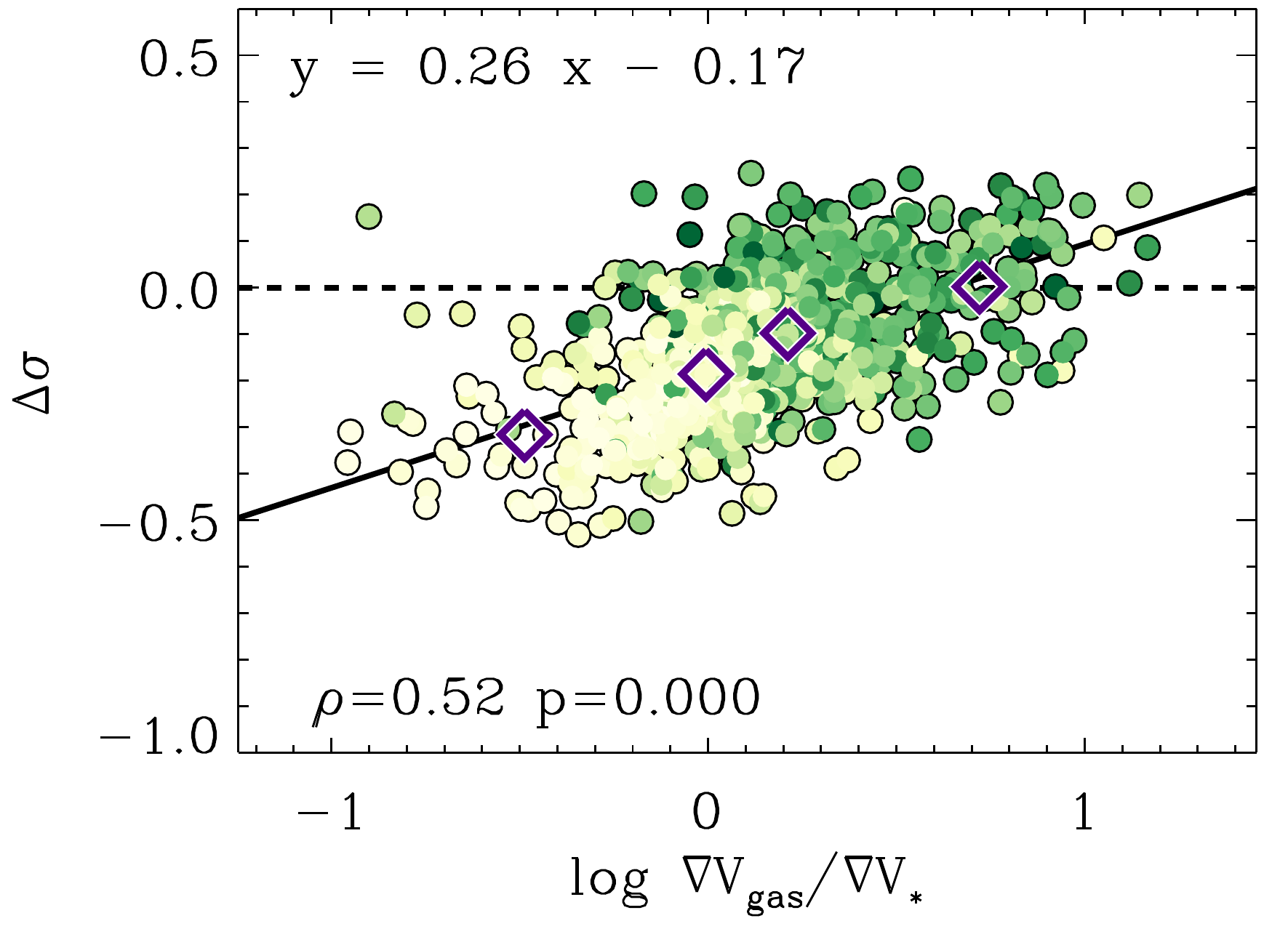}
\caption{The relation between $\Delta\sigma$ and log ($\nabla V_{\rm gas}/\nabla V_{*}$). We present a linear fit to the relation on the top and Spearman's coefficient $\rho$ and p-value at the bottom. The data are colour-coded by \bpt. Diamonds indicate the median $\Delta\sigma$ for each bin.}
\label{app1}
\end{figure}

Figure~\ref{app1} present a strong correlation between \dsig\ and log ($\nabla V_{\rm \lowercase{gas}}/\nabla V_{*}$) ($\rho$=0.52). We note that the conclusion of this study does not change when we use $\nabla V_{\rm \lowercase{gas}}/\nabla V_{*}$ as one of the key parameters instead of \rvg.

\section{Partial Least Squares regression}
\label{sec:pls}

We employ partial least squares (PLS; Wold 1966) regression, a multivariate data analysis technique, to statistically confirm the results found in the previous section. The PLS algorithm solves a regression model generalising a few latent variables (or principal components) that summarise the variance of independent variables and are also highly correlated with dependent variables. The model is 
\begin{align}\label{eq:pls} Y = X B + F \end{align}
where $X$ and $Y$ are, respectively, matrices for independent and dependent variables, $B$ is a matrix for PLS regression coefficients, and $F$ is an error matrix. The PLS algorithm decomposes the $X$ and $Y$ variables, maximising the covariance between latent variables $T$ and $U$ defined as
\begin{align} X = T P + E\\ Y = U Q + F \end{align}
where $T$ and $U$ are, respectively, linear combinations of $X$ and $Y$ with weights. $P$ and $Q$ are the orthogonal loading matrices, and $E$ and $F$ are the error matrices. The PLS algorithm assumes a linear relation between $T$ and $U$ to keep the correlation between $X$ and $Y$. 

To apply PLS regression to this study, we constructed the $X$ and $Y$ matrices containing, respectively, the values of nine galaxy parameters and the \dsig\ for 1090 galaxies. We utilised the PLSRegression function in the Python package Scikit-learn (Pedregosa et~al.\ 2012), which employs a non-linear iterative partial least squares (NIPALS; Lindgren et~al.\ 1993) algorithm for PLS regression on an incomplete data set. We determined the optimal number of PLS components (latent variables) using mean squared error (MSE) between observed and predicted $Y$ (\dsig\ in this study). MSE generally decreases when adding more components, but it converges after a certain number of components. We increased the number of PLS components until MSE converged, and selected the optimal number of PLS components. We estimated PLS regression coefficients $B$ and decomposed the variance in $Y$ (\dsig) into the contribution of each $X$ component (the nine galaxy parameters) using Equation~\ref{eq:pls}:
\begin{equation} {\rm Var}(Y) = \sum_{i=1}^{N=9} {\rm Var}(X_i B_i) + {\rm Var}(F) ~. \end{equation}

The proportion of $Y$ variance explained by each galaxy parameter has been calculated as the variance in ($X_i B_i$) divided by the variance in $Y$ (Table~\ref{tab:pls}). The PLS method finds that \rvg\ and \bpt\ are the two most important parameters, accounting for 76\% of the \dsig\ variance explained by the nine parameters. We also find a moderate contribution from \rre, accounting for around 10\% of the \dsig\ variance. The other galaxy parameters are less likely to have direct relevance to the variance in \dsig, explaining less than 5\% of the variance. The results from this PLS analysis support that \rvg, \bpt, and \rre\ are the key parameters in explaining the variance in \dsig, and that therefore they play an important role in producing the difference in the gas and stellar velocity dispersions. 

We applied the same analysis to \dsig$_\aper$ and found that \rvg\ is the parameter that most explains the variance in \dsig$_\aper$ (Table~\ref{tab:pls}). \bpt\ also show a moderate contribution to the \dsig$_\aper$ variance.

\begin{table*}\centering\caption{Proportion of the variance in \dsig\ and \dsig$_\aper$ explained by each galaxy parameter}\begin{tabular}{lrrrrrrrrr}
\hline\hline
 & log $M_\stars$/$M_{\odot}$& log R$_{\rm e}$& B/T & log Age & $[Z/H]$ & log SFR & \bpt & log (\rre) & log \rvg \\
\hline
\dsig & 0.00 & 0.05 & 0.02 & 0.01 & 0.01 & 0.05 & 0.35 & 0.10 & 0.41 \\
\dsig$_\aper$ & 0.04 & 0.04 & 0.02 & 0.00 & 0.00 & 0.07 & 0.18 & 0.02 & 0.61 \\
\hline\hline
\label{tab:pls}\end{tabular}
\end{table*}

%\bsp	% typesetting comment
\label{lastpage}

\begin{thebibliography}{}
\bibitem[\protect\citeauthoryear{Agertz et~al.}{2009}]{2009MNRAS.392..294A} Agertz O., Lake G., Teyssier R., Moore B., Mayer L., Romeo A.~B., 2009, MNRAS, 392, 294. doi:10.1111/j.1365-2966.2008.14043.x
\bibitem[\protect\citeauthoryear{Agertz, Teyssier, \& Moore}{2009}]{2009MNRAS.397L..64A} Agertz O., Teyssier R., Moore B., 2009, MNRAS, 397, L64. doi:10.1111/j.1745-3933.2009.00685.x
\bibitem[\protect\citeauthoryear{Aumer et~al.}{2010}]{2010ApJ...719.1230A} Aumer M., Burkert A., Johansson P.~H., Genzel R., 2010, ApJ, 719, 1230. doi:10.1088/0004-637X/719/2/1230
\bibitem[\protect\citeauthoryear{Avery et~al.}{2021}]{2021MNRAS.503.5134A} Avery C.~R., Wuyts S., F{\"o}rster Schreiber N.~M., Villforth C., Bertemes C., Chang W., Hamer S.~L., et~al., 2021, MNRAS, 503, 5134. doi:10.1093/mnras/stab780
\bibitem[\protect\citeauthoryear{Baldwin, Phillips, \& Terlevich}{1981}]{1981PASP...93....5B} Baldwin J.~A., Phillips M.~M., Terlevich R., 1981, PASP, 93, 5. doi:10.1086/130766
\bibitem[\protect\citeauthoryear{Barat et~al.}{2019}]{2019MNRAS.487.2924B} Barat D., D'Eugenio F., Colless M., Brough S., Catinella B., Cortese L., Croom S.~M., et~al., 2019, MNRAS, 487, 2924. doi:10.1093/mnras/stz1439
\bibitem[\protect\citeauthoryear{Barat et al.}{2020}]{2020MNRAS.498.5885B} Barat D., D'Eugenio F., Colless M., Sweet S.~M., Groves B., Cortese L., 2020, MNRAS, 498, 5885. doi:10.1093/mnras/staa2716
\bibitem[\protect\citeauthoryear{Barone et~al.}{2018}]{2018ApJ...856...64B} Barone T.~M., D'Eugenio F., Colless M., Scott N., van de Sande J., Bland-Hawthorn J., Brough S., et~al., 2018, ApJ, 856, 64. doi:10.3847/1538-4357/aaaf6e
\bibitem[\protect\citeauthoryear{Barone et~al.}{2020}]{2020ApJ...898...62B} Barone T.~M., D'Eugenio F., Colless M., Scott N., 2020, ApJ, 898, 62. doi:10.3847/1538-4357/ab9951
\bibitem[\protect\citeauthoryear{Barro et~al.}{2016}]{2016ApJ...820..120B} Barro G., Faber S.~M., Dekel A., Pacifici C., P{\'e}rez-Gonz{\'a}lez P.~G., Toloba E., Koo D.~C., et~al., 2016, ApJ, 820, 120. doi:10.3847/0004-637X/820/2/120
\bibitem[\protect\citeauthoryear{Barsanti et~al.}{2021}]{2021ApJ...911...21B} Barsanti S., Owers M.~S., McDermid R.~M., Bekki K., Bryant J.~J., Croom S.~M., Oh S., et~al., 2021, ApJ, 911, 21. doi:10.3847/1538-4357/abe5ac
\bibitem[\protect\citeauthoryear{Bassett et~al.}{2017}]{2017MNRAS.470.1991B} Bassett R., Bekki K., Cortese L., Couch W.~J., Sansom A.~E., van de Sande J., Bryant J.~J., et~al., 2017, MNRAS, 470, 1991. doi:10.1093/mnras/stx1000
\bibitem[\protect\citeauthoryear{Bekiaris et~al.}{2016}]{2016MNRAS.455..754B} Bekiaris G., Glazebrook K., Fluke C.~J., Abraham R., 2016, MNRAS, 455, 754. doi:10.1093/mnras/stv2292
\bibitem[Belfiore et al.(2015)]{2015MNRAS.449..867B} Belfiore, F., Maiolino, R., Bundy, K., et al.\ 2015, \mnras, 449, 867. doi:10.1093/mnras/stv296
\bibitem[Belfiore et al.(2016)]{2016MNRAS.461.3111B} Belfiore, F., Maiolino, R., Maraston, C., et al.\ 2016, \mnras, 461, 3111. doi:10.1093/mnras/stw1234
\bibitem[\protect\citeauthoryear{Bezanson et~al.}{2018}]{2018ApJ...868L..36B} Bezanson R., van der Wel A., Straatman C., Pacifici C., Wu P.-F., Bari{\v{s}}i{\'c} I., Bell E.~F., et~al., 2018, ApJL, 868, L36. doi:10.3847/2041-8213/aaf16b
\bibitem[\protect\citeauthoryear{Bland-Hawthorn et~al.}{2011}]{2011OExpr..19.2649B} Bland-Hawthorn J., Bryant J., Robertson G., Gillingham P., O'Byrne J., Cecil G., Haynes R., et~al., 2011, OExpr, 19, 2649. doi:10.1364/OE.19.002649
\bibitem[\protect\citeauthoryear{Bouch{\'e} et~al.}{2015}]{2015AJ....150...92B} Bouch{\'e} N., Carfantan H., Schroetter I., Michel-Dansac L., Contini T., 2015, AJ, 150, 92. doi:10.1088/0004-6256/150/3/92
\bibitem[\protect\citeauthoryear{Bryant et~al.}{2011}]{2011MNRAS.415.2173B} Bryant J.~J., O'Byrne J.~W., Bland-Hawthorn J., Leon-Saval S.~G., 2011, MNRAS, 415, 2173. doi:10.1111/j.1365-2966.2011.18841.x
\bibitem[\protect\citeauthoryear{Bryant et~al.}{2014}]{2014MNRAS.438..869B} Bryant J.~J., Bland-Hawthorn J., Fogarty L.~M.~R., Lawrence J.~S., Croom S.~M., 2014, MNRAS, 438, 869. doi:10.1093/mnras/stt2254
\bibitem[\protect\citeauthoryear{Bryant et~al.}{2015}]{2015MNRAS.447.2857B} Bryant J.~J., Owers M.~S., Robotham A.~S.~G., Croom S.~M., Driver S.~P., Drinkwater M.~J., Lorente N.~P.~F., et~al., 2015, MNRAS, 447, 2857. doi:10.1093/mnras/stu2635
\bibitem[\protect\citeauthoryear{Cappellari \& Emsellem}{2004}]{2004PASP..116..138C} Cappellari M., Emsellem E., 2004, PASP, 116, 138. doi:10.1086/381875
\bibitem[\protect\citeauthoryear{Cappellari}{2008}]{2008MNRAS.390...71C} Cappellari M., 2008, MNRAS, 390, 71. doi:10.1111/j.1365-2966.2008.13754.x
\bibitem[\protect\citeauthoryear{Cappellari}{2017}]{2017MNRAS.466..798C} Cappellari M., 2017, MNRAS, 466, 798. doi:10.1093/mnras/stw3020
\bibitem[\protect\citeauthoryear{Chen, Hao, \& Wang}{2008}]{2008ChJAA...8...25C} Chen X.-Y., Hao C.-N., Wang J., 2008, ChJAA, 8, 25. doi:10.1088/1009-9271/8/1/03
\bibitem[\protect\citeauthoryear{Cicone et~al.}{2014}]{2014A&A...562A..21C} Cicone C., Maiolino R., Sturm E., Graci{\'a}-Carpio J., Feruglio C., Neri R., Aalto S., et~al., 2014, A\&A, 562, A21. doi:10.1051/0004-6361/201322464
\bibitem[\protect\citeauthoryear{Colina, Arribas, \& Monreal-Ibero}{2005}]{2005ApJ...621..725C} Colina L., Arribas S., Monreal-Ibero A., 2005, ApJ, 621, 725. doi:10.1086/427683
\bibitem[\protect\citeauthoryear{Cortese et~al.}{2014}]{2014ApJ...795L..37C} Cortese L., Fogarty L.~M.~R., Ho I.-T., Bekki K., Bland-Hawthorn J., Colless M., Couch W., et~al., 2014, ApJL, 795, L37. doi:10.1088/2041-8205/795/2/L37
\bibitem[\protect\citeauthoryear{Croom et~al.}{2012}]{2012MNRAS.421..872C} Croom S.~M., Lawrence J.~S., Bland-Hawthorn J., Bryant J.~J., Fogarty L., Richards S., Goodwin M., et~al., 2012, MNRAS, 421, 872. doi:10.1111/j.1365-2966.2011.20365.x
\bibitem[\protect\citeauthoryear{Croom et~al.}{2021}]{2021MNRAS.505..991C} Croom S.~M., Owers M.~S., Scott N., Poetrodjojo H., Groves B., van de Sande J., Barone T.~M., et~al., 2021, MNRAS, 505, 991. doi:10.1093/mnras/stab229
\bibitem[\protect\citeauthoryear{D'Agostino et~al.}{2019}]{2019MNRAS.485L..38D} D'Agostino J.~J., Kewley L.~J., Groves B.~A., Medling A., Dopita M.~A., Thomas A.~D., 2019, MNRAS, 485, L38. doi:10.1093/mnrasl/slz028
\bibitem[\protect\citeauthoryear{D'Eugenio et~al.}{2021}]{2021MNRAS.504.5098D} D'Eugenio F., Colless M., Scott N., van der Wel A., Davies R.~L., van de Sande J., Sweet S.~M., et~al., 2021, MNRAS, 504, 5098. doi:10.1093/mnras/stab1146
\bibitem[\protect\citeauthoryear{de Jong et~al.}{2015}]{2015A&A...582A..62D} de Jong J.~T.~A., Verdoes Kleijn G.~A., Boxhoorn D.~R., Buddelmeijer H., Capaccioli M., Getman F., Grado A., et~al., 2015, A\&A, 582, A62. doi:10.1051/0004-6361/201526601
\bibitem[\protect\citeauthoryear{Debuhr, Quataert, \& Ma}{2012}]{2012MNRAS.420.2221D} Debuhr J., Quataert E., Ma C.-P., 2012, MNRAS, 420, 2221. doi:10.1111/j.1365-2966.2011.20187.x
\bibitem[\protect\citeauthoryear{Di Teodoro \& Fraternali}{2015}]{2015MNRAS.451.3021D} Di Teodoro E.~M., Fraternali F., 2015, MNRAS, 451, 3021. doi:10.1093/mnras/stv1213
\bibitem[\protect\citeauthoryear{Dib, Bell, \& Burkert}{2006}]{2006ApJ...638..797D} Dib S., Bell E., Burkert A., 2006, ApJ, 638, 797. doi:10.1086/498857
\bibitem[\protect\citeauthoryear{Dopita et~al.}{2013}]{2013ApJS..208...10D} Dopita M.~A., Sutherland R.~S., Nicholls D.~C., Kewley L.~J., Vogt F.~P.~A., 2013, ApJS, 208, 10. doi:10.1088/0067-0049/208/1/10
\bibitem[\protect\citeauthoryear{Driver et~al.}{2011}]{2011MNRAS.413..971D} Driver S.~P., Hill D.~T., Kelvin L.~S., Robotham A.~S.~G., Liske J., Norberg P., Baldry I.~K., et~al., 2011, MNRAS, 413, 971. doi:10.1111/j.1365-2966.2010.18188.x
\bibitem[\protect\citeauthoryear{Federrath et~al.}{2017}]{2017MNRAS.468.3965F} Federrath C., Salim D.~M., Medling A.~M., Davies R.~L., Yuan T., Bian F., Groves B.~A., et~al., 2017, MNRAS, 468, 3965. doi:10.1093/mnras/stx727
\bibitem[\protect\citeauthoryear{Gilhuly, Courteau, \& S{\'a}nchez}{2019}]{2019MNRAS.482.1427G} Gilhuly C., Courteau S., S{\'a}nchez S.~F., 2019, MNRAS, 482, 1427. doi:10.1093/mnras/sty2792
\bibitem[\protect\citeauthoryear{Glazebrook}{2013}]{2013PASA...30...56G} Glazebrook K., 2013, PASA, 30, e056. doi:10.1017/pasa.2013.34
\bibitem[\protect\citeauthoryear{Green et~al.}{2010}]{2010Natur.467..684G} Green A.~W., Glazebrook K., McGregor P.~J., Abraham R.~G., Poole G.~B., Damjanov I., McCarthy P.~J., et~al., 2010, Natur, 467, 684. doi:10.1038/nature09452
\bibitem[\protect\citeauthoryear{Greene \& Ho}{2005}]{2005ApJ...627..721G} Greene J.~E., Ho L.~C., 2005, ApJ, 627, 721. doi:10.1086/430590
\bibitem[\protect\citeauthoryear{Harborne et~al.}{2020}]{2020MNRAS.497.2018H} Harborne K.~E., van de Sande J., Cortese L., Power C., Robotham A.~S.~G., Lagos C.~D.~P., Croom S., 2020, MNRAS, 497, 2018. doi:10.1093/mnras/staa1847
\bibitem[\protect\citeauthoryear{Harrison et~al.}{2014}]{2014MNRAS.441.3306H} Harrison C.~M., Alexander D.~M., Mullaney J.~R., Swinbank A.~M., 2014, MNRAS, 441, 3306. doi:10.1093/mnras/stu515
\bibitem[\protect\citeauthoryear{Heckman}{2002}]{2002ASPC..254..292H} Heckman T.~M., 2002, ASPC, 254, 292
\bibitem[\protect\citeauthoryear{Ho}{2009}]{2009ApJ...699..638H} Ho L.~C., 2009, ApJ, 699, 638. doi:10.1088/0004-637X/699/1/638
\bibitem[\protect\citeauthoryear{Ho, Filippenko, \& Sargent}{1997}]{1997ApJS..112..315H} Ho L.~C., Filippenko A.~V., Sargent W.~L.~W., 1997, ApJS, 112, 315. doi:10.1086/313041
\bibitem[\protect\citeauthoryear{Ho et~al.}{2014}]{2014MNRAS.444.3894H} Ho I.-T., Kewley L.~J., Dopita M.~A., Medling A.~M., Allen J.~T., Bland-Hawthorn J., Bloom J.~V., et~al., 2014, MNRAS, 444, 3894. doi:10.1093/mnras/stu1653
\bibitem[\protect\citeauthoryear{Ho et~al.}{2016}]{2016Ap&SS.361..280H} Ho I.-T., Medling A.~M., Groves B., Rich J.~A., Rupke D.~S.~N., Hampton E., Kewley L.~J., et~al., 2016, Ap\&SS, 361, 280. doi:10.1007/s10509-016-2865-2
\bibitem[\protect\citeauthoryear{Johnson et~al.}{2018}]{2018MNRAS.474.5076J} Johnson H.~L., Harrison C.~M., Swinbank A.~M., Tiley A.~L., Stott J.~P., Bower R.~G., Smail I., et~al., 2018, MNRAS, 474, 5076. doi:10.1093/mnras/stx3016
\bibitem[\protect\citeauthoryear{Jones et~al.}{2016}]{2016ApJ...826...12J} Jones M.~L., Hickox R.~C., Black C.~S., Hainline K.~N., DiPompeo M.~A., Goulding A.~D., 2016, ApJ, 826, 12. doi:10.3847/0004-637X/826/1/12
\bibitem[\protect\citeauthoryear{Karouzos, Woo, \& Bae}{2016}]{2016ApJ...833..171K} Karouzos M., Woo J.-H., Bae H.-J., 2016, ApJ, 833, 171. doi:10.3847/1538-4357/833/2/171
\bibitem[\protect\citeauthoryear{Kauffmann et~al.}{2003}]{2003MNRAS.346.1055K} Kauffmann G., Heckman T.~M., Tremonti C., Brinchmann J., Charlot S., White S.~D.~M., Ridgway S.~E., et~al., 2003, MNRAS, 346, 1055. doi:10.1111/j.1365-2966.2003.07154.x
\bibitem[\protect\citeauthoryear{Kauffmann \& Heckman}{2009}]{2009MNRAS.397..135K} Kauffmann G., Heckman T.~M., 2009, MNRAS, 397, 135. doi:10.1111/j.1365-2966.2009.14960.x
\bibitem[\protect\citeauthoryear{Kewley et~al.}{2001}]{2001ApJ...556..121K} Kewley L.~J., Dopita M.~A., Sutherland R.~S., Heisler C.~A., Trevena J., 2001, ApJ, 556, 121. doi:10.1086/321545
\bibitem[\protect\citeauthoryear{Kewley et~al.}{2006}]{2006MNRAS.372..961K} Kewley L.~J., Groves B., Kauffmann G., Heckman T., 2006, MNRAS, 372, 961. doi:10.1111/j.1365-2966.2006.10859.x
\bibitem[Law et al.(2021)]{2021ApJ...915...35L} Law, D.~R., Ji, X., Belfiore, F., et al.\ 2021, \apj, 915, 35. doi:10.3847/1538-4357/abfe0a
\bibitem[\protect\citeauthoryear{Lindgren et~al.}{1993}]{} Lindgren F., Geladi P., Wold S., Journal of Chemometrics, 7 (1993), pp. 45-60. doi:10.1002/cem.1180070104
\bibitem[\protect\citeauthoryear{Medling et~al.}{2018}]{2018MNRAS.475.5194M} Medling A.~M., Cortese L., Croom S.~M., Green A.~W., Groves B., Hampton E., Ho I.-T., et~al., 2018, MNRAS, 475, 5194. doi:10.1093/mnras/sty127
\bibitem[\protect\citeauthoryear{Nelson \& Whittle}{1996}]{1996ApJ...465...96N} Nelson C.~H., Whittle M., 1996, ApJ, 465, 96. doi:10.1086/177405
\bibitem[\protect\citeauthoryear{Nesvadba et~al.}{2006}]{2006ApJ...650..693N} Nesvadba N.~P.~H., Lehnert M.~D., Eisenhauer F., Gilbert A., Tecza M., Abuter R., 2006, ApJ, 650, 693. doi:10.1086/507266
\bibitem[\protect\citeauthoryear{Oh et~al.}{2017}]{2017MNRAS.464.1466O} Oh K., Schawinski K., Koss M., Trakhtenbrot B., Lamperti I., Ricci C., Mushotzky R., et~al., 2017, MNRAS, 464, 1466. doi:10.1093/mnras/stw2467
\bibitem[\protect\citeauthoryear{Oh et~al.}{2019}]{2019ApJ...880..112O} Oh K., Ueda Y., Akiyama M., Suh H., Koss M.~J., Kashino D., Hasinger G., 2019, ApJ, 880, 112. doi:10.3847/1538-4357/ab288b
\bibitem[\protect\citeauthoryear{Orr et~al.}{2020}]{2020MNRAS.496.1620O} Orr M.~E., Hayward C.~C., Medling A.~M., Gurvich A.~B., Hopkins P.~F., Murray N., Pineda J.~L., et~al., 2020, MNRAS, 496, 1620. doi:10.1093/mnras/staa1619
\bibitem[\protect\citeauthoryear{Owers et~al.}{2017}]{2017MNRAS.468.1824O} Owers M.~S., Allen J.~T., Baldry I., Bryant J.~J., Cecil G.~N., Cortese L., Croom S.~M., et~al., 2017, MNRAS, 468, 1824. doi:10.1093/mnras/stx562
\bibitem[\protect\citeauthoryear{Owers et~al.}{2019}]{2019ApJ...873...52O} Owers M.~S., Hudson M.~J., Oman K.~A., Bland-Hawthorn J., Brough S., Bryant J.~J., Cortese L., et~al., 2019, ApJ, 873, 52. doi:10.3847/1538-4357/ab0201
\bibitem[\protect\citeauthoryear{Pedregosa et~al.}{2012}]{2012arXiv1201.0490P} Pedregosa F., Varoquaux G., Gramfort A., Michel V., Thirion B., Grisel O., Blondel M., et~al., 2012, arXiv, arXiv:1201.0490
\bibitem[\protect\citeauthoryear{Rich, Kewley, \& Dopita}{2015}]{2015ApJS..221...28R} Rich J.~A., Kewley L.~J., Dopita M.~A., 2015, ApJS, 221, 28. doi:10.1088/0067-0049/221/2/28
\bibitem[\protect\citeauthoryear{Robotham et~al.}{2017}]{2017MNRAS.466.1513R} Robotham A.~S.~G., Taranu D.~S., Tobar R., Moffett A., Driver S.~P., 2017, MNRAS, 466, 1513. doi:10.1093/mnras/stw3039
\bibitem[\protect\citeauthoryear{Rupke, Veilleux, \& Sanders}{2005}]{2005ApJ...632..751R} Rupke D.~S., Veilleux S., Sanders D.~B., 2005, ApJ, 632, 751. doi:10.1086/444451
\bibitem[\protect\citeauthoryear{Rupke, Veilleux, \& Sanders}{2005}]{2005ApJS..160..115R} Rupke D.~S., Veilleux S., Sanders D.~B., 2005, ApJS, 160, 115. doi:10.1086/432889
\bibitem[\protect\citeauthoryear{Scott et~al.}{2017}]{2017MNRAS.472.2833S} Scott N., Brough S., Croom S.~M., Davies R.~L., van de Sande J., Allen J.~T., Bland-Hawthorn J., et~al., 2017, MNRAS, 472, 2833. doi:10.1093/mnras/stx2166
\bibitem[\protect\citeauthoryear{Scott et~al.}{2018}]{2018MNRAS.481.2299S} Scott N., van de Sande J., Croom S.~M., Groves B., Owers M.~S., Poetrodjojo H., D'Eugenio F., et~al., 2018, MNRAS, 481, 2299. doi:10.1093/mnras/sty2355
\bibitem[\protect\citeauthoryear{Shanks et~al.}{2013}]{2013Msngr.154...38S} Shanks T., Belokurov V., Chehade B., Croom S.~M., Findlay J.~R., Gonzalez-Solares E., Irwin M.~J., et~al., 2013, Msngr, 154, 38
\bibitem[\protect\citeauthoryear{Shapiro et~al.}{2010}]{2010MNRAS.402.2140S} Shapiro K.~L., Falc{\'o}n-Barroso J., van de Ven G., de Zeeuw P.~T., Sarzi M., Bacon R., Bolatto A., et~al., 2010, MNRAS, 402, 2140. doi:10.1111/j.1365-2966.2009.16111.x
\bibitem[\protect\citeauthoryear{Sharp et~al.}{2006}]{2006SPIE.6269E..0GS} Sharp R., Saunders W., Smith G., Churilov V., Correll D., Dawson J., Farrel T., et~al., 2006, SPIE, 6269, 62690G. doi:10.1117/12.671022
\bibitem[\protect\citeauthoryear{Shetty et~al.}{2020}]{2020ApJ...901..101S} Shetty S., Bershady M.~A., Westfall K.~B., Cappellari M., Drory N., Law D.~R., Yan R., et~al., 2020, ApJ, 901, 101. doi:10.3847/1538-4357/ab9b8e
\bibitem[\protect\citeauthoryear{Shih, Stockton, \& Kewley}{2013}]{2013ApJ...772..138S} Shih H.-Y., Stockton A., Kewley L., 2013, ApJ, 772, 138. doi:10.1088/0004-637X/772/2/138
\bibitem[\protect\citeauthoryear{Steidel et~al.}{2010}]{2010ApJ...717..289S} Steidel C.~C., Erb D.~K., Shapley A.~E., Pettini M., Reddy N., Bogosavljevi{\'c} M., Rudie G.~C., et~al., 2010, ApJ, 717, 289. doi:10.1088/0004-637X/717/1/289
\bibitem[\protect\citeauthoryear{Sutherland \& Dopita}{1993}]{1993ApJS...88..253S} Sutherland R.~S., Dopita M.~A., 1993, ApJS, 88, 253. doi:10.1086/191823
\bibitem[\protect\citeauthoryear{S{\'a}nchez-Bl{\'a}zquez et~al.}{2006}]{2006MNRAS.371..703S} S{\'a}nchez-Bl{\'a}zquez P., Peletier R.~F., Jim{\'e}nez-Vicente J., Cardiel N., Cenarro A.~J., Falc{\'o}n-Barroso J., Gorgas J., et~al., 2006, MNRAS, 371, 703. doi:10.1111/j.1365-2966.2006.10699.x
\bibitem[\protect\citeauthoryear{Sharp \& Bland-Hawthorn}{2010}]{2010ApJ...711..818S} Sharp R.~G., Bland-Hawthorn J., 2010, ApJ, 711, 818. doi:10.1088/0004-637X/711/2/818
\bibitem[\protect\citeauthoryear{Tamburro et~al.}{2009}]{2009AJ....137.4424T} Tamburro D., Rix H.-W., Leroy A.~K., Mac Low M.-M., Walter F., Kennicutt R.~C., Brinks E., et~al., 2009, AJ, 137, 4424. doi:10.1088/0004-6256/137/5/4424
\bibitem[\protect\citeauthoryear{Taylor et~al.}{2011}]{2011MNRAS.418.1587T} Taylor E.~N., Hopkins A.~M., Baldry I.~K., Brown M.~J.~I., Driver S.~P., Kelvin L.~S., Hill D.~T., et~al., 2011, MNRAS, 418, 1587. doi:10.1111/j.1365-2966.2011.19536.x
\bibitem[\protect\citeauthoryear{Thomas et~al.}{2018}]{2018ApJ...861L...2T} Thomas A.~D., Kewley L.~J., Dopita M.~A., Groves B.~A., Hopkins A.~M., Sutherland R.~S., 2018, ApJL, 861, L2. doi:10.3847/2041-8213/aacce7
\bibitem[\protect\citeauthoryear{van de Sande et~al.}{2017}]{2017ApJ...835..104V} van de Sande J., Bland-Hawthorn J., Fogarty L.~M.~R., Cortese L., d'Eugenio F., Croom S.~M., Scott N., et~al., 2017, ApJ, 835, 104. doi:10.3847/1538-4357/835/1/104
\bibitem[\protect\citeauthoryear{Varidel et~al.}{2016}]{2016PASA...33....6V} Varidel M., Pracy M., Croom S., Owers M.~S., Sadler E., 2016, PASA, 33, e006. doi:10.1017/pasa.2016.3
\bibitem[\protect\citeauthoryear{Varidel et~al.}{2019}]{2019MNRAS.485.4024V} Varidel M.~R., Croom S.~M., Lewis G.~F., Brewer B.~J., Di Teodoro E.~M., Bland-Hawthorn J., Bryant J.~J., et~al., 2019, MNRAS, 485, 4024. doi:10.1093/mnras/stz670
\bibitem[\protect\citeauthoryear{Varidel et~al.}{2020}]{2020MNRAS.495.2265V} Varidel M.~R., Croom S.~M., Lewis G.~F., Fisher D.~B., Glazebrook K., Catinella B., Cortese L., et~al., 2020, MNRAS, 495, 2265. doi:10.1093/mnras/staa1272
\bibitem[\protect\citeauthoryear{Veilleux, Cecil, \& Bland-Hawthorn}{2005}]{2005ARA&A..43..769V} Veilleux S., Cecil G., Bland-Hawthorn J., 2005, ARA\&A, 43, 769. doi:10.1146/annurev.astro.43.072103.150610
\bibitem[\protect\citeauthoryear{Weiner et~al.}{2006}]{2006ApJ...653.1049W} Weiner B.~J., Willmer C.~N.~A., Faber S.~M., Harker J., Kassin S.~A., Phillips A.~C., Melbourne J., et~al., 2006, ApJ, 653, 1049. doi:10.1086/508922
\bibitem[\protect\citeauthoryear{Weiner et~al.}{2009}]{2009ApJ...692..187W} Weiner B.~J., Coil A.~L., Prochaska J.~X., Newman J.~A., Cooper M.~C., Bundy K., Conselice C.~J., et~al., 2009, ApJ, 692, 187. doi:10.1088/0004-637X/692/1/187
\bibitem[\protect\citeauthoryear{Wisnioski et~al.}{2011}]{2011MNRAS.417.2601W} Wisnioski E., Glazebrook K., Blake C., Wyder T., Martin 
C., Poole G.~B., Sharp R., et~al., 2011, MNRAS, 417, 2601. doi:10.1111/j.1365-2966.2011.19429.x
\bibitem[\protect\citeauthoryear{Wold}{1966}]{} Wold H., P.R. Krishnaiah (Ed.), Multivariate Analysis, Academic Press, New York (1966), pp. 391-420
\bibitem[\protect\citeauthoryear{Woo, Son, \& Bae}{2017}]{2017ApJ...839..120W} Woo J.-H., Son D., Bae H.-J., 2017, ApJ, 839, 120. doi:10.3847/1538-4357/aa6894
\bibitem[\protect\citeauthoryear{York et~al.}{2000}]{2000AJ....120.1579Y} York D.~G., Adelman J., Anderson J.~E., Anderson S.~F., Annis J., Bahcall N.~A., Bakken J.~A., et~al., 2000, AJ, 120, 1579. doi:10.1086/301513
\bibitem[\protect\citeauthoryear{Zhou et~al.}{2017}]{2017MNRAS.470.4573Z} Zhou L., Federrath C., Yuan T., Bian F., Medling A.~M., Shi Y., Bland-Hawthorn J., et~al., 2017, MNRAS, 470, 4573. doi:10.1093/mnras/stx1504
\end{thebibliography}
\end{document}